\documentclass[10pt,eqsecnum,twocolumn]{revtex4}
\usepackage[colorlinks=true]{hyperref}
\usepackage{graphicx} 
\usepackage{color} 
\usepackage{amsthm}
\usepackage{amsmath}
\usepackage{amssymb}
\usepackage{bbm,amsfonts}
\usepackage{setspace}
\usepackage{nomencl}
\usepackage{amstext} 
\usepackage{subfigure}
\usepackage{comment}

\newcommand{\Heff}{H_\text{eff}}

\begin{document}

\title{Effective Hamiltonian models of the cross-resonance gate}
\begin{abstract}
Effective Hamiltonian methods are utilized to model the two-qubit cross-resonance gate for both the ideal two-qubit case and when higher levels are included. Analytic expressions are obtained in the qubit case and the higher-level model is solved both perturbatively and numerically with the solutions agreeing well in the weak drive limit. The methods are applied to parameters from recent experiments and accounting for classical cross-talk effects results in good agreement between theory and experimental results.
\end{abstract}

\author{Easwar Magesan}
\affiliation{IBM T.J. Watson Research Center, Yorktown Heights, NY 10598, USA}
\author{Jay M. Gambetta}
\affiliation{IBM T.J. Watson Research Center, Yorktown Heights, NY 10598, USA}

\maketitle

\section{Introduction}

Implementing high-fidelity quantum operations is a central problem in the field of experimental quantum information processing (QIP). Building a universal fault-tolerant quantum computer requires the ability to perform a high-fidelity two quantum-bit (qubit) entangling gate and, while many proposals have been put forth for implementing two-qubit gates, few experimental demonstrations have achieved fidelities near those determined by fault-tolerant analyses~\cite{Dennis2002,Harrington2007}. The \emph{cross-resonance (CR)} gate~\cite{Paraoanu,Rigetti10} has recently been utilized in superconducting circuit systems~\cite{Gambetta2017} to achieve a two-qubit CNOT gate with high fidelity exceeding 0.99~\cite{Sheldon2016} and has been used in small-scale multi-qubit demonstrations of fault-tolerant protocols~\cite{Chow2014,Corcoles2015,Takita2016,Takita2017}. Here we provide a theoretical analysis of the CR gate and outline effective Hamiltonian methods that can be used to obtain a description of the gate dynamics. The techniques developed are directly applicable in realistic systems as shown by the improved experimental calibration and high two-qubit gate fidelity of Ref.~\cite{Sheldon2016}. 

Broadly speaking, the goal of effective Hamiltonian theory is to model some set of complex dynamics on a large system via a more compact Hamiltonian on a set of smaller subsystems or subspaces. Effective Hamiltonian methods have been utilized in various areas of physics and chemistry including nuclear, atomic-molecular, optical, and condensed matter systems. In particular, techniques for the adiabatic elimination of higher energy levels in a system have been widely studied, common examples including the Schrieffer-Wolff transformation~\cite{SW,Bravyi11} and Born-Oppenheimer approximation~\cite{BornOpp}. Here we restrict attention to effective Hamiltonian constructions based on unitary (canonical) operations that transform the Hamiltonian $H$ on the full Hilbert space $\mathcal{H}$ into a block-diagonal Hamiltonian $\Heff$ with the two-block case corresponding to the standard Schrieffer-Wolff transformation.

We outline the perturbative construction of an effective Hamiltonian where the desired block-diagonal form is enforced at each order. The advantage of the perturbative construction lies in obtaining analytical expressions for components of $\Heff$ that hold in the weak perturbation limit. The second construction is based on an exact multi-block-diagonalization technique~\cite{Cederbaum} which finds the block-diagonal Hamiltonian that is closest to the true Hamiltonian under the principle of least action. This method has the advantage of being valid in the strong drive regime however it is not possible to compute general analytic expressions for the Hamiltonian components except in simple cases such as the basic two-qubit model. These methods are applied to obtain an effective model for the CR gate Hamiltonian in the two-qubit model as well as when higher levels are included. As a concrete example we use the parameters of Ref.~\cite{Sheldon2016} and find good agreement between the perturbative and exact multi-block diagonalization approaches in the weak-drive limit with higher levels included. However a discrepancy is found between the theory and experiment since Ref.~\cite{Sheldon2016} finds the presence of extra unwanted terms in the Hamiltonian. We propose classical cross-talk between the two transmons from the CR drive as a potential source for this discrepancy and extend the analysis to include this effect. After doing so we find very good agreement between the theoretical predictions and experimental results of Ref.~\cite{Sheldon2016}. 

\smallskip

The paper is structured as follows. First in Sec.~\ref{sec:Initial} we describe the starting Hamiltonian for the analysis which consists of two transmons dispersively coupled to a resonator. We outline a method to find an effective block-diagonal Hamiltonian for the two transmon-resonator system which when projected onto the zero-excitation subspace of the resonator provides an effective Hamiltonian for the two transmon system alone. Next in Sec.~\ref{sec:Qubitmodel} we model the transmons as ideal qubits and find an analytic expression for the effective CR Hamiltonian under the principle of least action. In Sec.~\ref{sec:higherlevel} we model the transmons as Duffing oscillators and find perturbative expressions that hold in the weak-drive limit. We also perform a numerical analysis of the exact block-diagonalization technique using the parameters of Ref.~\cite{Sheldon2016}. In Sec.~\ref{sec:Targetdrive} we analyze classical cross-talk occurring from the CR drive to model the results of Ref.~\cite{Sheldon2016}. For clarity of the presentation the mathematical methods and details of effective Hamiltonian theory are contained in the appendix (Sec.~\ref{sec:EffectiveHam}) with the main text focused mainly on discussion of the application of the methods to the CR gate.

\section{Initial Hamiltonian and effective two-transmon Hamiltonian} \label{sec:Initial}

We start with a Hamiltonian describing the standard cQED~\cite{BHW04} set-up of two transmons~\cite{Koch07}, modeled as Duffing oscillators, coupled to a bus resonator
\begin{gather}
H_\text{sys}=\sum_{j=1}^2\left(\bar{\omega}_jb_j^{\dagger}b_j + \frac{\delta_1}{2}b_j^{\dagger}b_j(b_j^{\dagger}b_j-\openone)\right) + \omega_r c^\dagger c \nonumber \\
 + \sum_{j=1}^2g_j(b_j^\dagger c + b_jc^{\dagger}),
\end{gather}
where we set $\hbar = 1$. Here $\bar{\omega}_jb_j^{\dagger}b_j + \frac{\delta_j}{2}b_j^{\dagger}b_j(b_j^{\dagger}b_j-\openone)$ is the Duffing Hamiltonian of the $j$'th transmon ($j=1,2$) with $\bar{\omega}_j$ and $\delta_j$ being the $01$ transition frequency and anharmonicity of the $j$'th transmon respectively. The resonator Hamiltonian $\omega_r c^\dagger c$ is a single-mode harmonic oscillator with fundamental frequency $\omega_r \gg \bar{\omega}_{1(2)}$. Each transmon is coupled to the resonator by a Jaynes-Cummings Hamiltonian with coupling strength $g_j$ and since we are in the dispersive regime,
\begin{align}
\left|\frac{g_j}{\Delta_{j,r}}\right| & \ll 1,
\end{align}
where $\Delta_{j,r} = \bar{\omega}_j - \omega_r$ is the detuning of the $j$'th transmon to the resonator. Note the total number of excitations is a symmetry of the system as the total excitation operator commutes with $H_\text{sys}$.

 Let us write $H_\text{sys}$ as the sum of two Hamiltonians $H_{\text{sys},0}$ and $H_{\text{sys},1}$
\begin{align}
H_{\text{sys},0} &= \sum_{j=1}^2\left(\bar{\omega}_jb_j^{\dagger}b_j + \frac{\delta_1}{2}b_j^{\dagger}b_j(b_j^{\dagger}b_j-\openone)\right) + \omega_r c^\dagger c, \nonumber \\
H_{\text{sys},1} &=  \sum_{j=1}^2g_j(b_j^\dagger c + b_jc^{\dagger}),
\end{align}
where $H_{\text{sys},0}$ is diagonal and $H_{\text{sys},1}$ contains all of the coupling terms so is off-diagonal. Moving into the frame rotating at $\omega_r$ via the unitary
\begin{align}
R&= e^{-it\omega_r\left(b_1^\dagger b_1 + b_2^\dagger b_2 + c^\dagger c\right)},
\end{align}
gives
\begin{align}
H_\text{sys}&=\sum_{j=1}^2\left(\Delta_{j,r} b_j^{\dagger}b_j + \frac{\delta_j}{2}b_j^{\dagger}b_j(b_j^{\dagger}b_j-\openone) \right) \nonumber \\
&+ \sum_{j=1}^2g_j(b_j^\dagger c + b_jc^{\dagger}),
\end{align}
where $\Delta_{j,r} = \bar{\omega}_j-\omega_r$. In this frame the diagonal part of $H_\text{sys}$ is independent of the resonator photon number. Therefore $H_\text{sys}$ is the direct sum of infinitely many identical copies (blocks) where each copy corresponds to a photon number $\{0,1,2,...\}$,
\begin{align}
	\text{diag}(H_\text{sys}) &=
	\left[ \begin{array}{cccccc}
	0_p	&	\:	&	\: 	&	\:	&	\:	&	\:	\\
	\:	&	1_p	&	 \:	&	\:	&	\:	&	\:	\\
	\:	&	\:	&	 2_p	&	\:	&	\:	&	\:	\\
	\:	&	\:	&	 \:	&	\:.	&	\:	&	\:	\\
	\:	&	\:	&	 \:	&	\:	&	\:.	&	\:	\\
	\:	&	\:	&	 \:	&	\:	&	\:	&	\:.	\\
	\end{array} \right].
\end{align}
All of the photon number blocks $j_p$ describe the same two-transmon Hamiltonian and each block can be broken into sub-blocks labeled by excitation number of the transmons
\begin{align}
	j_p &=
	\left[ \begin{array}{cccccc}
	0	&	\:	&	\: 	&	\:	&	\:	&	\:	\\
	\:	&	\{\Delta_{i,r}\}	&	 \:	&	\:	&	\:	&	\:	\\
	\:	&	\:	&	\{2\Delta_{i,r}+\delta_i, \Delta_{1,r} + \Delta_{2,r}\}	&	\:	&	\:	&	\:	\\
	\:	&	\:	&	 \:	&	\:.	&	\:	&	\:	\\
	\:	&	\:	&	 \:	&	\:	&	\:.	&	\:	\\
	\:	&	\:	&	 \:	&	\:	&	\:	&	\:.	\\
	\end{array} \right].
\end{align}
Hence we can denote every possible excitation block by the label $(j_P,k_T)$ where ``P" refers to photon and ``T" refers to transmon. 

Now, since the photon number blocks $j_p$ support the same Hamiltonian, the blocks $(m_P,k_T)$ and $(r_P,k_T)$ have the exact same form for $m \neq r$ and we can group all of the levels with the same transmon excitation number into a single infinite-dimensional block. The first block corresponds to zero excitations in the transmons
\begin{gather}
\{(0_P,0_T), (1_P,0_T),....,(m_P,0_T),...\} \nonumber \\
= \{|0_P\rangle \otimes |00\rangle, |1_P\rangle \otimes |00\rangle,....,|m_P\rangle \otimes |00\rangle,...\},
\end{gather}
at 0 energy scale, the second block corresponds to one transmon excitation
\begin{gather}
\{(0_P,1_T),....,(m_P,1_T),...\} \nonumber \\
=  \{|0_P\rangle \otimes |01\rangle, |0_P\rangle \otimes |10\rangle,....,|m_P\rangle \otimes |01\rangle,|m_P\rangle \otimes |10\rangle,...\},
\end{gather}
at the energy scale $\{\Delta_{1,r},\Delta_{2,r}\}$, and the third block corresponds to two transmon excitations
\begin{align}
& \{(0_P,2_T),....,(m_P,2_T),...\}\nonumber \\
&=  \{|0_P\rangle \otimes |02\rangle, |0_P\rangle \otimes |11\rangle, |0_P\rangle \otimes |20\rangle,....,\nonumber \\
& \: \: \: \: \: \: |m_P\rangle \otimes |02\rangle, |m_P\rangle \otimes |11\rangle, |m_P\rangle \otimes |20\rangle,...\},
\end{align}
at an energy scale of $\{2\Delta_{1,r}+\delta_1, \Delta_{1,r} + \Delta_{2,r}, 2\Delta_{2,r}+\delta_2\}$. The fourth block will correspond to three transmon excitations and four energies of the same order, and in general the $k$'th block will correspond to $k-1$ transmon excitations and $k$ different energies of the same order. 

Hence $\text{diag}(H_\text{sys})$ is written as
\begin{align}
	\text{diag}(H_\text{sys}) &=
	\left[ \begin{array}{cccccc}
	0_T	&	\:	&	\: 	&	\:	&	\:	&	\:	\\
	\:	&	1_T	&	 \:	&	\:	&	\:	&	\:	\\
	\:	&	\:	&	 2_T	&	\:	&	\:	&	\:	\\
	\:	&	\:	&	 \:	&	\:.	&	\:	&	\:	\\
	\:	&	\:	&	 \:	&	\:	&	\:.	&	\:	\\
	\:	&	\:	&	 \:	&	\:	&	\:	&	\:.	\\
	\end{array} \right],
\end{align}
where each block $k_T$ has energy approximately on the order of $k\Delta_{j,r}$. Since the coupling terms preserve total excitation number there are \emph{no coupling terms connecting elements within each block}. All coupling terms connect different blocks which are detuned on the order of $\Delta_{j,r}$, that is there are only couplings between the blocks $((j+1)_P,(k-1)_T)$ and $((j-1)_P,(k+1)_T)$. Since these blocks are detuned on the order of $\Delta_{j,r}$, which is assumed to be much larger than the coupling strengths $g_j$, the couplings can be adiabatically eliminated to give an effective block-diagonal Hamiltonian for the whole system as outlined via the methods in Sec.~\ref{sec:EffectiveHam}. In the dispersive regime where $\omega_r$ is much larger than the transmon frequencies one can obtain an effective Hamiltonian for the two transmons by projecting onto the zero-excitation subspace of the bus which gives
\begin{gather}
H_\text{sys}^{(0)}=\sum_{j=1}^2\left(\tilde{\omega}_jb_j^{\dagger}b_j + \frac{\delta_j}{2}b_j^{\dagger}b_j(b_j^{\dagger}b_j-1)\right) \nonumber \\
+ J(b_1^\dagger b_2 + b_1b_2^{\dagger}),\label{eq:Hsys0}
\end{gather} 
where $\tilde{\omega}_1$ and $\tilde{\omega}_2$ are the dressed qubit frequencies and to lowest order the exchange coupling is given by
\begin{align}
J&=\frac{g_1g_2(\bar{\omega}_1+\bar{\omega}_2 - 2\omega_r)}{2(\bar{\omega}_1-\omega_r)(\bar{\omega}_2-\omega_r)}.
\end{align}
The general drive Hamiltonian is modeled as 
\begin{align}\label{eq:DriveHam}
H_\text{d} &=\sum_{j=1}^2\left[\Omega_{X_j}(t)\cos(\omega_{d_j}t) + \Omega_{Y_j}(t)\sin(\omega_{d_j}t)\right](b_j^{\dagger} + b_j),
\end{align}
where $\omega_{d_j}$ is the drive frequency on transmon $j$ and $\Omega_{X_j}(t)$, $\Omega_{Y_j}(t)$ are the drive amplitudes on the quadratures of transmon $j$. For now we will focus mainly on the case of only a drive term on the $X$ quadrature of qubit 1 (control) which gives a total Hamiltonian
\begin{gather}
H_\text{T}= H_\text{sys}^{(0)}+H_\text{d} \nonumber \\
=\sum_{j=1}^2\left(\tilde{\omega}_jb_j^{\dagger}b_j + \frac{\delta_j}{2}b_j^{\dagger}b_j(b_j^{\dagger}b_j-1) \right) + J(b_1^\dagger b_2 + b_1b_2^{\dagger})\nonumber \\
+\sum_{j=1}^2\left[\Omega_{X_j}(t)\cos(\omega_{d_j}t) + \Omega_{Y_j}(t)\sin(\omega_{d_j}t)\right](b_j^{\dagger} + b_j).\label{eq:TotalHam}
\end{gather}
We take $H_\text{T}$ to form the basis of our analysis and analyze the ideal qubit model next.

\section{Effective CR Hamiltonian for a qubit model}\label{sec:Qubitmodel}

In the qubit model the anharmonicity is infinite so the qubit subspace is perfectly isolated and $H_\text{T}$ is given by
\begin{gather}
H_\text{T} = \sum_{j=1}^2 \tilde{\omega}_jb_j^{\dagger}b_j  + J(b_1^\dagger b_2 + b_1b_2^{\dagger})  + \Omega(t)\cos(\tilde{\omega}_2 t)(b_1^{\dagger} + b_1),
\end{gather}
where $b_j$ is a two-level operator, the control qubit is driven at the frequency of the target qubit, and for simplicity we assume $\Omega (t) = \Omega$ is a constant amplitude drive on the $X$ quadrature of the control qubit only. We derive an exact expression for the full qubit CR Hamiltonian using the method of Ref.~\cite{Cederbaum} and find an effective $ZX$ term that agrees with expressions derived previously using alternative methods~\cite{Rigetti10}. First, we move into the frame rotating at $\tilde{\omega}_2$ on both qubits and make the RWA by ignoring fast-rotating terms. Writing $\cos(\tilde{\omega}_2 t) = (e^{-i\tilde{\omega}_2 t} + e^{i\tilde{\omega}_2 t})/2$, defining
\begin{align}
R=e^{-i\tilde{\omega}_2\left(b_1^\dagger b_1 + b_2^\dagger b_2\right)t},
\end{align}
and ignoring fast-rotating terms gives the Hamiltonian
\begin{align}
H_R &= R^\dagger H_\text{T} R -i R^\dagger \dot{R} \nonumber \\
&= \Delta b_1^{\dagger}b_1 + J(b_1^\dagger b_2 + b_1b_2^{\dagger}) + \frac{\Omega}{2}\left(b_1^\dagger + b_1\right) \nonumber \\
&=
	\left[ \begin{array}{cccc}
	0			&	0				&	 \Omega/2		&	0\\
	0			&	0				&	 J				&	 \Omega/2\\
	 \Omega/2	&	J				&	 \Delta			&	0\\
	0			&	 \Omega/2		&	 0				&	\Delta\\
	\end{array} \right],
\end{align}
where $\Delta =\tilde{\omega}_1-\tilde{\omega}_2$. From the form of $H_R$ there are naturally two $2\times 2$ blocks, one corresponding to the states $|00\rangle$, $|01\rangle$ with energy scale $0$ and the other corresponding to the states $|10\rangle$, $|11\rangle$ with energy scale $\Delta$. Using the method of Sec.~\ref{sec:Leastaction}, which in this case corresponds to the standard Schrieffer-Wolff transformation, one can find the closest block-diagonal Hermitian matrix to $H_R$ under the principle of least action. Let $X$ be the eigenvector matrix of $H_R$, that is, $X$ has columns consisting of the normalized eigenvectors of $H_R$. Let $\overline{X}$ be the unnormalized version of $X$ with columns given by
\begin{align} 
\left[ \begin{array}{c}
	\frac{ \left(J^2 + \sqrt{J^2(J^2+\Omega^2)}\right) \left(\Delta + \sqrt{2J^2 + \Delta^2+\Omega^2-2 \sqrt{J^2(J^2+\Omega^2)}}\right)}{J\Omega^2}\\
		-\frac{ \left(\Delta + \sqrt{2J^2 + \Delta^2+\Omega^2-2 \sqrt{J^2(J^2+\Omega^2)}}\right)}{\Omega}	\\
	 \frac{ \left(J^2 - \sqrt{J^2(J^2+\Omega^2)}\right)}{J\Omega}\\
	1\\
	\end{array} \right], \nonumber  \\
\left[ \begin{array}{c}
	\frac{ \left(J^2 - \sqrt{J^2(J^2+\Omega^2)}\right) \left(\Delta + \sqrt{2J^2 + \Delta^2+\Omega^2+2 \sqrt{J^2(J^2+\Omega^2)}}\right)}{J\Omega^2}\\
		-\frac{ \left(\Delta + \sqrt{2J^2 + \Delta^2+\Omega^2+2 \sqrt{J^2(J^2+\Omega^2)}}\right)}{\Omega}	\\
	 \frac{ \left(J^2 + \sqrt{J^2(J^2+\Omega^2)}\right)}{J\Omega}\\
	1\\
	\end{array} \right], \nonumber  \\
\left[ \begin{array}{c}
	\frac{ -\left(J^2 + \sqrt{J^2(J^2+\Omega^2)}\right) \left(-\Delta + \sqrt{2J^2 + \Delta^2+\Omega^2-2 \sqrt{J^2(J^2+\Omega^2)}}\right)}{J\Omega^2}\\
		\frac{ \left(-\Delta + \sqrt{2J^2 + \Delta^2+\Omega^2-2 \sqrt{J^2(J^2+\Omega^2)}}\right)}{\Omega}	\\
	 \frac{ \left(J^2 - \sqrt{J^2(J^2+\Omega^2)}\right)}{J\Omega}\\
	1\\
	\end{array} \right], \nonumber  \\
\left[ \begin{array}{c}
	\frac{ \left(-J^2 + \sqrt{J^2(J^2+\Omega^2)}\right) \left(-\Delta + \sqrt{2J^2 + \Delta^2+\Omega^2+2 \sqrt{J^2(J^2+\Omega^2)}}\right)}{J\Omega^2}\\
		\frac{ \left(-\Delta + \sqrt{2J^2 + \Delta^2+\Omega^2+2 \sqrt{J^2(J^2+\Omega^2)}}\right)}{\Omega}	\\
	 \frac{ \left(J^2 + \sqrt{J^2(J^2+\Omega^2)}\right)}{J\Omega}\\
	1\\
	\end{array} \right].\nonumber 
\end{align}
Approximating 
\begin{align}
\frac{ \left(J^2 \pm \sqrt{J^2(J^2+\Omega^2)}\right)}{J\Omega} \sim 1,
\end{align}
and re-scaling the eigenvectors implies $\overline{X}$ takes the form
\begin{align}
\left[ \begin{array}{c}
	1\\
	-1\\
	- \frac{ \Omega }{(\Delta + \sqrt{J^2 + \Delta^2+(\Omega-J)^2})}\\
	\frac{ \Omega }{(\Delta + \sqrt{J^2 + \Delta^2+(\Omega-J)^2})}\\\\
	\end{array} \right],
\left[ \begin{array}{c}
	1\\
	1\\
	- \frac{ \Omega }{(\Delta + \sqrt{J^2 + \Delta^2+(\Omega+J)^2})}\\
	-\frac{ \Omega }{(\Delta + \sqrt{J^2 + \Delta^2+(\Omega+J)^2})}\\\\
	\end{array} \right],
\end{align}
\begin{align}
\left[ \begin{array}{c}
	1\\
	-1\\
	 \frac{ \Omega }{(-\Delta + \sqrt{J^2 + \Delta^2+(\Omega-J)^2})}\\
	-\frac{ \Omega }{(-\Delta + \sqrt{J^2 + \Delta^2+(\Omega-J)^2})}\\\\
	\end{array} \right],
\left[ \begin{array}{c}
	1\\
	1\\
	\frac{ \Omega }{(-\Delta + \sqrt{J^2 + \Delta^2+(\Omega+J)^2})}\\
	\frac{ \Omega }{(-\Delta + \sqrt{J^2 + \Delta^2+(\Omega+J)^2})}\\\\
	\end{array} \right].
\end{align}
The least-action unitary $T$ that block-diagonalizes $H_R$ is given by
\begin{align}
T&=X X_{BD}^\dagger X_P^{-\frac{1}{2}},
\end{align}
where $X_{BD}$ is the block-diagonalization of $X$ and $X_P= X_{BD} X_{BD}^\dagger$. We have
\begin{align}
\left(\overline{X}_P\right)^{-\frac{1}{2}}(1:2,1:2) &= 
\left[ \begin{array}{cc}
	\frac{1}{\sqrt{2}} & 0 \\
	0 & \frac{1}{\sqrt{2}} \\
	\end{array} \right],
\end{align}
and $\left(\overline{X}_P\right)^{-\frac{1}{2}}(3:4,3:4)$ is the $2\times 2$ matrix
\begin{align}
\left[ \begin{array}{cc}
a & b \\
b & a \\
\end{array} \right],
\end{align}
where
\begin{align}
a&= \frac{ -2\Delta + \sqrt{J^2 + \Delta^2 + (\Omega-J)^2} +  \sqrt{J^2 + \Delta^2 + (\Omega+J)^2}}{2\sqrt{2}\Omega} \nonumber \\
b&= \frac{ - \sqrt{J^2 + \Delta^2 + (\Omega-J)^2} +  \sqrt{J^2 + \Delta^2 + (\Omega+J)^2}}{2\sqrt{2}\Omega}.
\end{align}
Ignoring terms of order $J^2$ gives the following unnormalized columns for $T$,
\begin{align}
\left[ \begin{array}{c}
	\sqrt{2}\\
	0\\
	\frac{ -\sqrt{2}\Omega }{(\Delta + \sqrt{ \Delta^2+ \Omega^2})}\\
	0\\\\
	\end{array} \right],
\left[ \begin{array}{c}
	0\\
	\sqrt{2}\\
	0\\
	\frac{-\sqrt{2}\Omega }{(\Delta + \sqrt{ \Delta^2+ \Omega^2})}\\\\
	\end{array} \right],\nonumber \\
\left[ \begin{array}{c}
	\sqrt{2}\\
	0\\
	\frac{-\sqrt{2}\Omega}{( \Delta - \sqrt{\Delta^2 + \Omega^2})}\\
	0\\\\
	\end{array} \right],
\left[ \begin{array}{c}
	0\\
	\sqrt{2}\\
	0\\
	\frac{-\sqrt{2}\Omega}{( \Delta - \sqrt{\Delta^2 + \Omega^2})}\\\\
	\end{array} \right].
\end{align}
Finally computing the block-diagonal of $H_R$ and moving back to the physical frame consisting of the transmons rotating at their respective frequencies gives the block-diagonal Hamiltonian $H_\text{CR}$ with $2\times 2$ blocks given by
\begin{align}
\frac{1}{2}\left[ \begin{array}{cc}
	\Delta-\sqrt{\Delta^2 + \Omega^2} &  -\frac{J \Omega}{\sqrt{\Delta^2 + \Omega^2}} \\
	 -\frac{J \Omega}{\sqrt{\Delta^2 + \Omega^2}} & \Delta-\sqrt{\Delta^2 + \Omega^2}  \\
	\end{array} \right].
\end{align}
\begin{align}
\frac{1}{2}\left[ \begin{array}{cc}
	-\Delta+\sqrt{\Delta^2 + \Omega^2} & \frac{J \Omega}{\sqrt{\Delta^2 + \Omega^2}} \\
	\frac{J \Omega}{\sqrt{\Delta^2+ \Omega^2}} & -\Delta+\sqrt{\Delta^2 + \Omega^2}\\\\
	\end{array} \right].
\end{align}
The $ZX$ term is thus given by
\begin{align}
\text{tr}\left(H_\text{CR} \left[\frac{ZX}{2}\right]\right) &= -\frac{J \Omega}{\sqrt{\Delta^2 + \Omega^2}},
\end{align}
where by virtue of the system Hamiltonian definition, the two-qubit Pauli operators are scaled by $\frac{1}{2}$ (in an $n$-qubit system they are scaled by $\frac{1}{2^{n-1}}$). The Stark-shift term on the control qubit is given by
\begin{align}
\text{tr}\left(H_\text{CR} \left[\frac{ZI}{2}\right]\right) &= \Delta-\sqrt{\Delta^2 + \Omega^2},
\end{align}
and so in total
\begin{align}
H_\text{CR} &= \left(\Delta-\sqrt{\Delta^2 + \Omega^2}\right) \frac{Z\mathbbm{1}}{2} -\left(\frac{J \Omega}{\sqrt{\Delta^2 + \Omega^2}}\right)\frac{ZX}{2}.
\end{align}

\section{Effective CR Hamiltonian for a higher-level model}\label{sec:higherlevel}

For a model including higher levels the approach is to first dress $H_\text{sys}^{(0)}$ in Eq.~\ref{eq:Hsys0} and then rotate the drive term into this frame. The system is then moved into the frame rotating at the target qubit frequency on both qubits and an RWA is performed. In this rotating frame the control $|0\rangle$ and $|1\rangle$ states define two subspaces that are far detuned by $\sim \Delta$ and an effective block-diagonal Hamiltonian is obtained via the perturbative analysis of Sec.~\ref{sec:EffectivePerturbation}. Unlike the qubit case, exact analytical expressions are not straightforward to obtain and so realistic parameters are used for the exact method of Sec.~\ref{sec:Leastaction}. For these parameters we find that the perturbative expressions and the exact block-diagonalization agree up to medium power drives of $\Omega \sim 50$ MHz with the exact method holding for much larger values of $\Omega$.

To start we assume that $\frac{J}{|\tilde{\omega}_1-\tilde{\omega}_2|} \ll 1$ and obtain an effective diagonal Hamiltonian for $H_\text{sys}^{(0)}$. Letting $U$ be the diagonalizing (dressing) unitary the effective diagonal Hamiltonian is given by
\begin{align}
\tilde{H}_\text{sys}^{(0)} = U^\dagger H_\text{sys}^{(0)} U,
\end{align}
where to second order in the two-qubit subspace
\begin{align}
\tilde{H}_\text{sys}^{(0)} &= \omega_1 \frac{Z\openone}{2} + \omega_2 \frac{\openone Z}{2} + \xi\frac{ZZ}{2},
\end{align}
with
\begin{align}
\omega_1 &= -\tilde{\omega}_1 - \frac{J^2}{\Delta} - \xi, \\
\omega_2 &= -\tilde{\omega}_2 + \frac{J^2}{\Delta} - \xi, \\
\xi &= -\frac{J^2(\delta_1+\delta_2)}{(\Delta + \delta_1)(\delta_2-\Delta)}.\label{eq:ZZ}
\end{align}
The presence of higher levels has produced an effective $ZZ$ interaction in the two-qubit subspace. The drive term of Eq.~\ref{eq:DriveHam} is rotated into this frame by applying the diagonalizing unitary $U$,
\begin{align}\label{eq:Hddiag}
\tilde{H}_\text{d} &= \sum_{j=1}^2\left[\Omega_{X_j}(t)\cos(\omega_{d_j}t) + \Omega_{Y_j}(t)\sin(\omega_{d_j}t)\right] \tilde{B}_j,
\end{align}
where $\tilde{B}_j = U^\dagger (b_j^{\dagger} + b_j)U$ for $j=1,2$.
We set $\omega_{d_1}=\omega_{d_2}=\omega_d$ and the Hamiltonian in the dressed frame is given by
\begin{align}
H(t)=\tilde{H}_\text{sys}^{(0)} + \tilde{H}_\text{d}(t),
\end{align}
Moving into the frame rotating at $\omega_d$ on both transmons and making the RWA as outlined in Sec.~\ref{sec:RWADuffing} gives the Hamiltonian
\begin{align}
H_\text{RWA}&=\tilde{H}_\text{drift}+\tilde{H}_\text{d,RWA},
\end{align}
where
\begin{align}
\tilde{H}_\text{drift}&:=\tilde{H}_\text{sys}^{(0)}-\tilde{H}_A, \nonumber \\
\tilde{H}_\text{d,RWA}&:= (R^\dagger \tilde{H}_d R)^\text{RWA}, \nonumber \\
\tilde{H}_A &= \omega_d(b_1^\dagger b_1 +  b_2^\dagger b_2),
\end{align}
and the matrix elements of $ (R^\dagger H_\text{d,diag}R)^\text{RWA}$ are given by the cases in Eq.~\ref{eq:Cases}. The drive frequency on the control transmon, $\omega_d$, is set to be the average of the dressed target transmon frequencies over the ground and excited states of the control transmon,
\begin{align}
\omega_d &= \frac{\tilde{H}_\text{sys}^{(0)}(11)-\tilde{H}_\text{sys}^{(0)}(10) + \tilde{H}_\text{sys}^{(0)}(01)-\tilde{H}_\text{sys}^{(0)}(00)}{2}.\label{eq:omegad}
\end{align}

We suppose the states are ladder-ordered as $\{00,01,10,11,02,20,03,12,21,30,....,0d,...,d0\}$ with $F$ denoting the permutation matrix that moves to ladder ordering from standard Kronecker ordering. To second order in $J$ the $\{00,01\}$ subspace has energy $\frac{J^2}{\Delta}$, the $\{10,11\}$ subspace has energy $\Delta + \frac{J^2}{\Delta}$, and $\text{\{rest\}}$ is assumed to be detuned from both of these subspaces. Loosely speaking, the energy of the state $|jk\rangle$ is given by 
\begin{align}
j\Delta + \frac{j(j-1)}{2}\delta_1 + \frac{k(k-1)}{2}\delta_2,
\end{align}
so that $H_\text{drift}$ is naturally partitioned according to the relative detunings with respect to $\omega_d$. Therefore the space can be partitioned as $\{00,01\},\{10,11\},\{\text{rest}\}$. The off-diagonal elements have a magnitude set by $\Omega  \left(\frac{J}{\Delta}\right)^m$ for $m \geq 0$. Let us now analyze the perturbative approach to obtain analytic expressions in the weak-drive limit and then investigate the exact method under the principle of least action.

\subsection{Effective perturbative Hamiltonian}\label{sec:higherlevelpert}

Under the assumption $\frac{\Omega}{\Delta} \ll 1$ a canonical transformation can be perturbatively constructed to find an effective block-diagonal Hamiltonian via the method outlined in Sec~\ref{sec:Blockdiagonalization}. We assume the drive term in Eq.~\ref{eq:Hddiag} contains only a drive on the $X$ quadrature of the control with a constant amplitude $\Omega$. The unperturbed Hamiltonian, denoted $H_0$, can be defined in a few different ways. For instance it can be defined via the block-diagonals of $H_{\text{RWA}}$,
\begin{align}
H_0&=P_{0001}H_\text{RWA}P_{0001} + P_{1011}H_\text{RWA}P_{1011} \nonumber \\
&+ P_{\text{rest}}H_\text{RWA}P_{\text{rest}},
\end{align}
with perturbative term given by
\begin{align}
H_1&=\frac{H_\text{RWA}-H_0}{\Omega},
\end{align}
so that
\begin{align}
H_\text{RWA} &= H_0 + \lambda H_1.
\end{align}

Unfortunately, defining $H_0$ to be block-diagonal does not provide simple analytic expressions for the effective block-diagonal Hamiltonian components because one needs to analytically compute the inverse of $H_0$ (see Sec.~\ref{sec:Blockdiagonalization}). As a result, we approach the construction by defining an unperturbed Hamiltonian via the diagonals of $H_{\text{RWA}}$
\begin{align}
H_0&= \text{diag}(H_{\text{RWA}}),
\end{align}
and define the perturbative term by
\begin{align}
H_1&=\frac{H_\text{RWA}-H_0}{\Omega}.
\end{align}
At each order we enforce block-diagonality as usual where the diagonal unperturbed Hamiltonian is treated as block-diagonal. In this picture all terms of the Hamiltonian containing the drive are included in the perturbation Hamiltonian and the inverse of $H_0$ is simple to compute. The order parameter is given by $\lambda=\Omega$ and
\begin{align}
H_\text{RWA} &= H_0 + \lambda H_1.
\end{align}

The perturbation proceeds as follows. The effective Hamiltonian takes the form
\begin{align}
\Heff &=\sum_{m=0}^\infty \lambda^m H^{(m)},
\end{align}
where $H^{(0)} = H_0$ is diagonal (block-diagonal) and for $m > 0$,
\begin{align}
H^{(m)} &= i\left[S^{(m)},H_0\right] + H_x^{(m)},
\end{align}
 with $H_x^{(m)}$ defined in Sec.~\ref{sec:Blockdiagonalization}. We define the $m$'th order approximation to $\Heff$ by
\begin{align}
H_\text{eff}^{(m)} &=H^{(0)}+\lambda H^{(1)} + \lambda^2 H^{(2)} + ... + \lambda^m H^{(m)}.
\end{align}
where at each order $H^{(m)}$ is enforced to be block-diagonal by the choice of $S^{(m)}$. Usually for $m=1$, $H_x^{(1)} = H_1$ is orthogonal to the desired form of $\Heff^{(m)}$ and so $S^{(1)}$ typically eliminates the first-order term $H^{(1)}$. Here however by choosing $H_0$ to be diagonal, $H_1$ has non-zero super/sub-diagonals of order $\frac{J}{\Delta}\Omega$. Therefore while we assume as usual that $S^{(1)}$ is off-block-diagonal, the non-zero super/sub-diagonals of $H_1$ survive to give a contribution at first order in $\Omega$ so that $H_\text{eff}^{(1)}$ has leading diagonals of order $ \frac{J}{\Delta}\Omega$ and unchanged diagonal elements (which have shifts of order $\frac{J^2}{\Delta}$ from the dressing) .

Keeping terms to first order in $J$,
\begin{align}
H_x^{(2)} &= -\frac{1}{2}\left[S^{(1)},\left[S^{(1)},H_0\right]\right] + i\left[S^{(1)},H_1\right]
\end{align}
has sub/super diagonals equal to 0 and contributions of order $\Omega^2$ on the diagonals. Thus the second order term contributes only to the diagonals and the off-diagonals of $H_\text{eff}^{(2)}$ are the same as in the first-order expression,
\begin{align}
H_\text{eff}^{(1)}[1,2] = H_\text{eff}^{(2)}[1,2] = -\frac{J\Omega_{X,1}}{2\Delta}, \nonumber \\
H_\text{eff}^{(1)}[3,4] = H_\text{eff}^{(2)}[3,4] = -\frac{J\Omega_{X,1}(\Delta-\delta_1)}{2\Delta(\Delta + \delta_1)}.
\end{align}
Going to third-order one again obtains corrections to the off-diagonals and we use the third-order effective Hamiltonian for the analytic expressions of the Hamiltonian. Moving back into the physical frame to restore the correct energies relative to the respective qubit frequencies gives the final Hamiltonian $H_\text{CR}$,
 \begin{gather}
H_\text{CR} = H_\text{eff} + \left(\omega_d - \omega_{d_1}\right)F(b^\dagger b \otimes \openone)F^\dagger.
\end{gather}
where
\begin{align}
\omega_{d_1} &= \frac{\tilde{H}_\text{sys}^{(0)}(11)-\tilde{H}_\text{sys}^{(0)}(01) + \tilde{H}_\text{sys}^{(0)}(10)-\tilde{H}_\text{sys}^{(0)}(00)}{2}
\end{align}
is the dressed frequency of the control qubit.

The ZX coefficient to third order is given by
\begin{align}
\frac{ZX}{2}_\text{coeff} &= \frac{ZX}{2}_\text{linear} \nonumber \\
& + \frac{ J\Omega^3 \delta_1^2 (3 \delta_1^3 + 11 \delta_1^2 \Delta + 15 \delta_1 \Delta^2 + 9 \Delta^3)
}{ 2\Delta^3 (\delta1 + \Delta)^3 (\delta_1 + 2 \Delta)(3 \delta_1 + 2 \Delta)},
\end{align}
where
\begin{align}
\frac{ZX}{2}_\text{linear}  &= -\frac{J\Omega}{\Delta}\left(\frac{\delta_1}{\delta_1+\Delta}\right),
\end{align}
and the full set of Pauli coefficients is given in Sec.~\ref{sec:Paulicoeff}. The poles in the $ZX$ expression occur at $\Delta = 0$, $-\frac{\delta_1}{2}$, $-\delta_1$, $-\frac{3\delta_1}{2}$.  The point $\Delta = 0$ corresponds to the qubits on-resonance and the point $\Delta=-\delta_1$ corresponds to the $\omega_{01}^{(1)}=\omega_{12}^{(2)}$. The points $\Delta=-\frac{\delta_1}{2}$ and $\Delta=-\delta_1$, $-\frac{3\delta_1}{2}$ are two-photon processes, the first of which corresponds to $\omega_{01}^{(1)}=\frac{\omega_{02}^{(2)}}{2}$. If these points are avoided one expects the perturbative expressions to model the system well in the weak drive limit. 

\subsection{Effective Hamiltonian from principle of least action}\label{sec:PrincipleLA}

\begin{figure}[h!]
\centering
\includegraphics[width=0.43\textwidth]{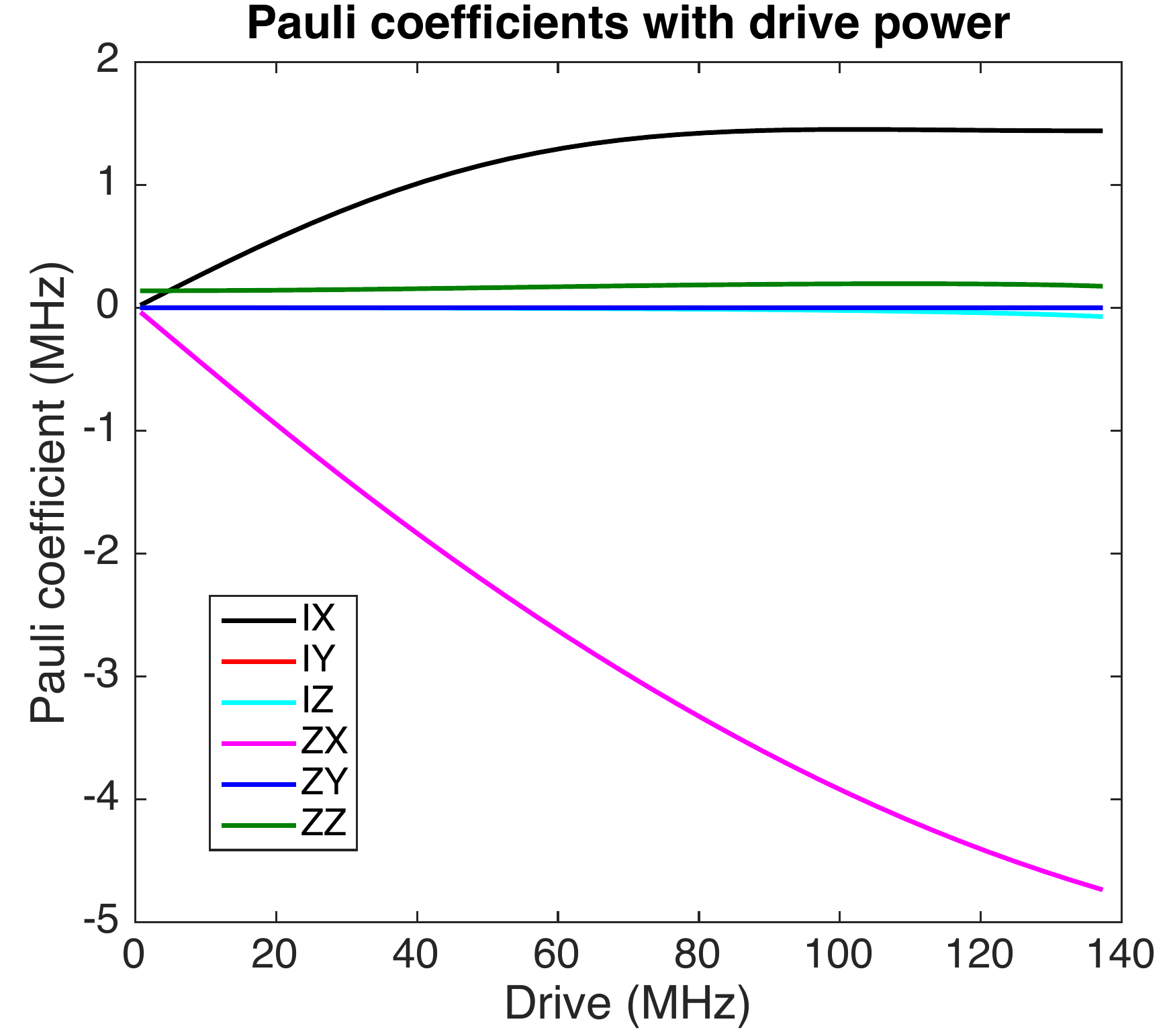}
\caption{\label{fig:Allcoeffs_Sarah}  Pauli coefficients for model with higher levels.}
\end{figure}

\begin{figure}[h!]
\centering
\includegraphics[width=0.43\textwidth]{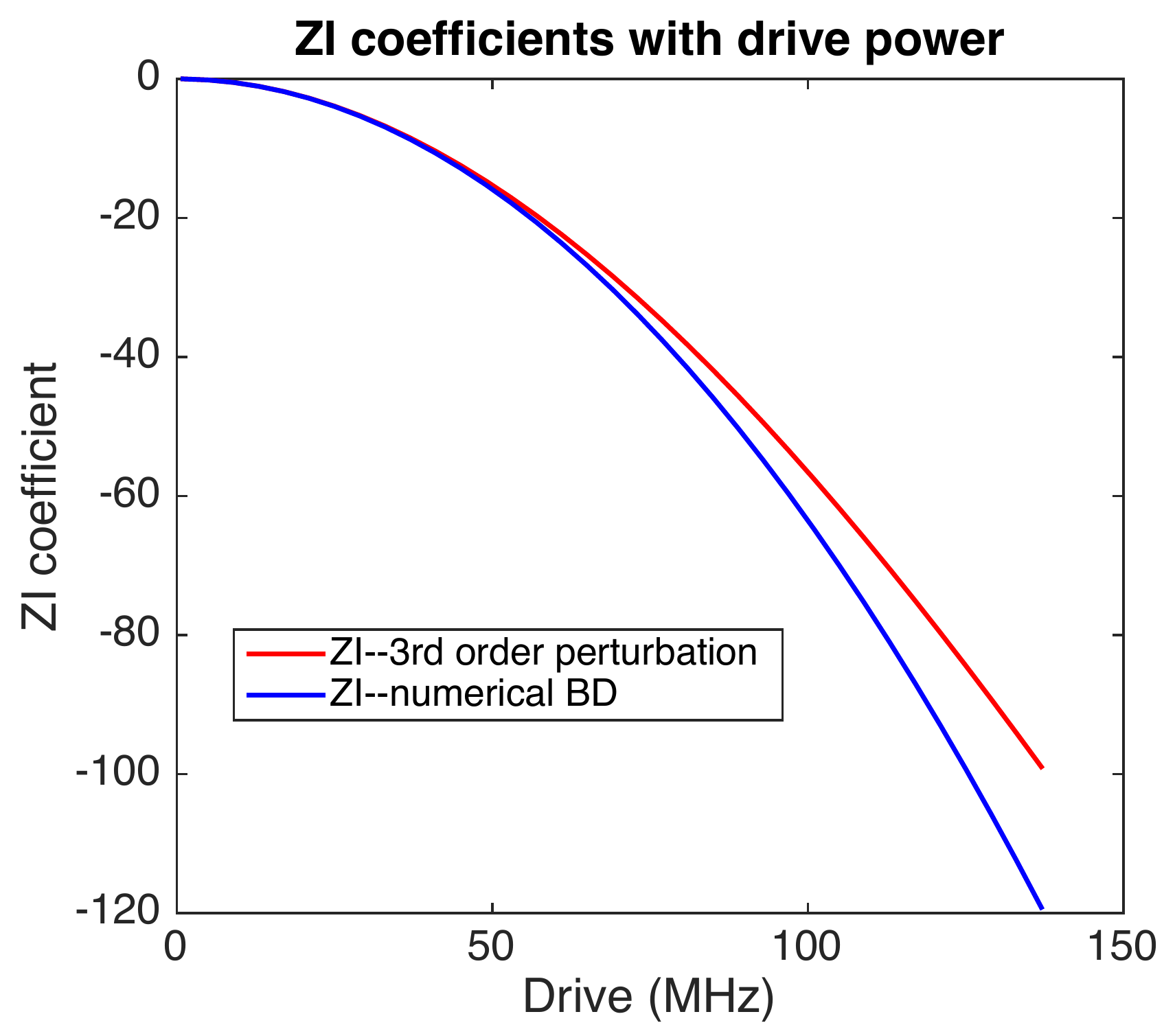}
\caption{\label{fig:ZI}  ZI Pauli coefficient for model with higher levels.}
\end{figure}

\begin{figure}[h!]
\centering
\includegraphics[width=0.43\textwidth]{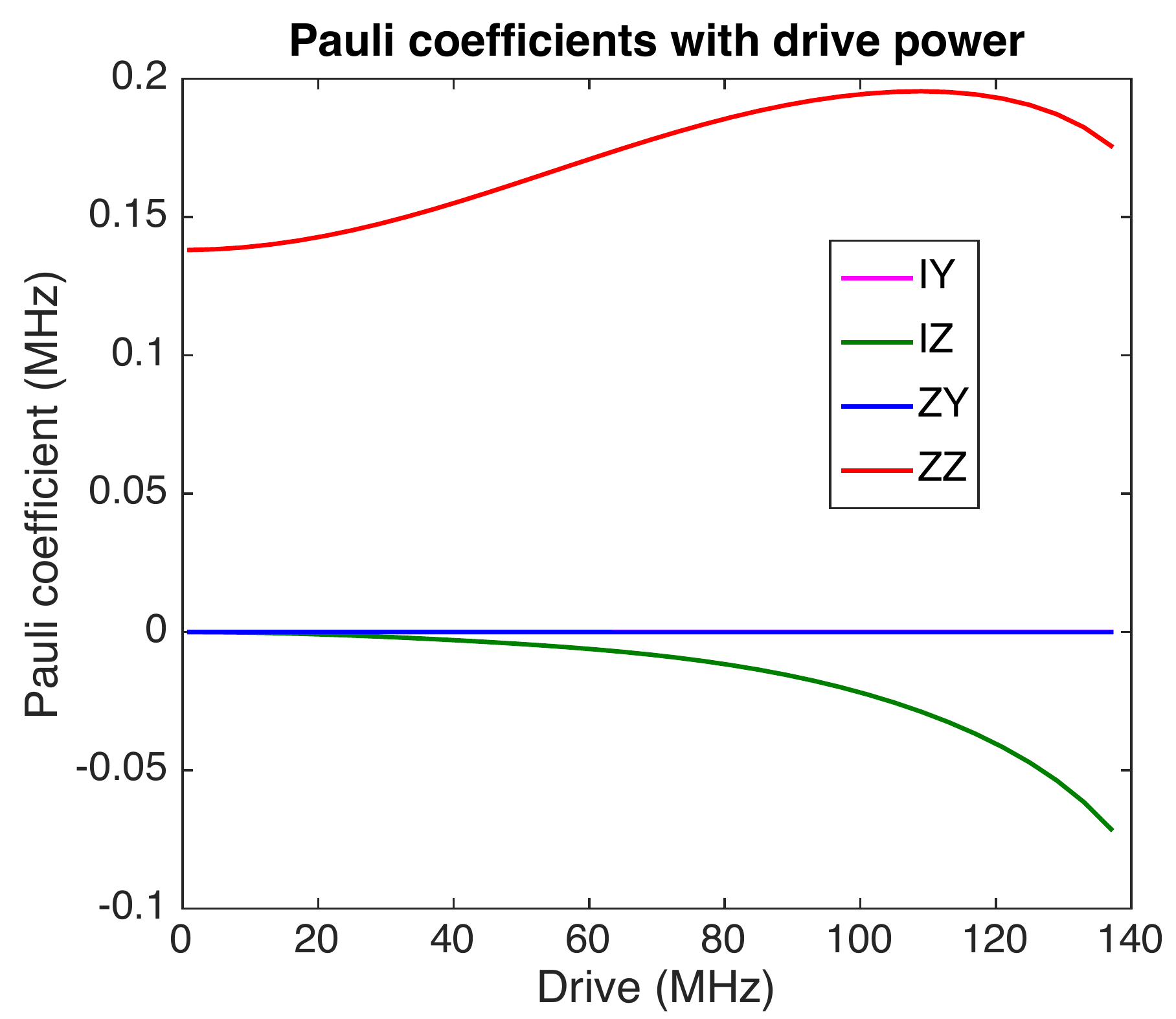}
\caption{\label{fig:IZ_ZY_IY_ZZ}  IY, IZ, ZY, and ZZ Pauli coefficients for model with higher levels.}
\end{figure}

An effective block-diagonal CR Hamiltonian obtained under the principle of least action (outlined in Sec.~\ref{sec:Leastaction}) provides a valid model in the limit of strong drives where the perturbative model breaks down. Since a general analytic expression for the effective Hamiltonian can not be obtained we use the device parameters of Ref.~\cite{Sheldon2016} to form the basis of our study; $\omega_1/2\pi = 5.114  \: \text{GHz}$, $\omega_2/2\pi = 4.914 \: \text{GHz}$, $\delta_ 1/2\pi = -0.330 \: \text{GHz}$, $\delta_2/2\pi = -0.330\: \text{GHz}$, $g_1/2\pi = 0.098 \: \text{GHz}$, $g_2/2\pi = 0.083 \: \text{GHz}$, $\omega_r/2\pi = 6.31 \: \text{GHz}$, and $\xi/2\pi = 277 \: \text{kHz}$. Using the approximation from Eq.\ref{eq:ZZ}
\begin{align}
\xi &= -\frac{2J^2(\delta_1+\delta_2)}{(\Delta_{12}+\delta_1)(\delta_2-\Delta_{12})},
\end{align}
the exchange coupling rate is given by $J/2\pi= 3.8 \: \text{MHz}$. Fig.~\ref{fig:Allcoeffs_Sarah} contains all of the relevant Pauli coefficients except $ZI$ which is given in Fig.~\ref{fig:ZI} and diverges quickly since the control qubit is driven far off-resonance. The presence of higher-levels and finite anharmonicity produces a large $IX$ term in the Hamiltonian that is not present in the pure qubit model. The $ZX$ and $IX$ coefficients have the largest magnitude and so the other coefficients are also contained alone in Fig.~\ref{fig:IZ_ZY_IY_ZZ}. The $IZ$ and $ZZ$ terms do not deviate significantly from their initial values as the drive amplitude increases. Note that the non-zero offset of the $ZZ$ coefficient corresponds to the static $ZZ$ term.

\begin{figure}[h!]
\centering
\includegraphics[width=0.43\textwidth]{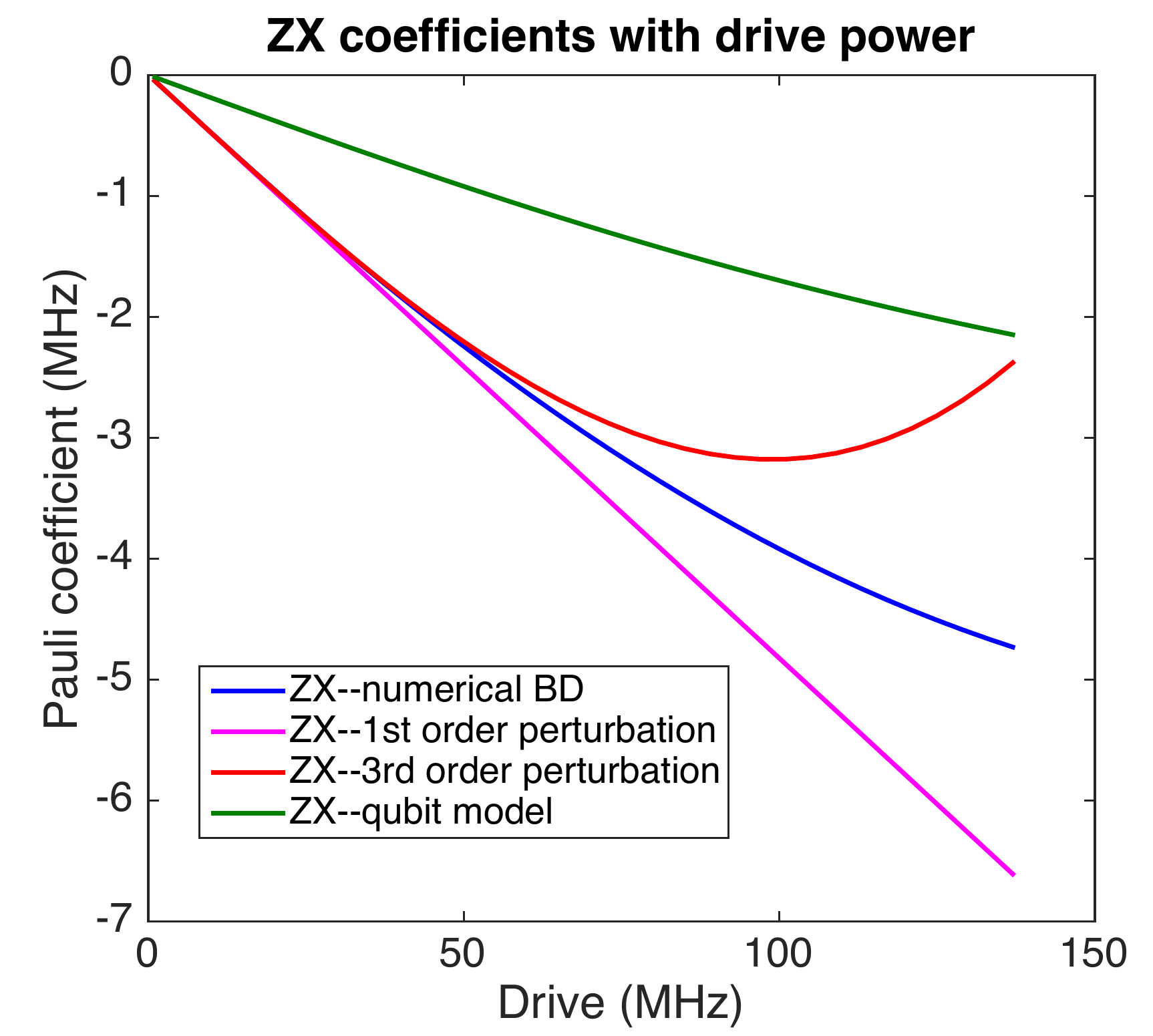}
\caption{\label{fig:ZX_Sarah}  Various expressions for $ZX$ term.}
\end{figure}

Importantly, there is no $IY$ term present which is also expected from the perturbative expressions for the Pauli coefficients in Sec.~\ref{sec:Paulicoeff}. This is in contrast to the experimental results of Ref.~\cite{Sheldon2016} where there is a large $IY$ component for this for set of parameters. We revisit this discrepancy in Sec.~\ref{sec:Targetdrive}.
Fig.~\ref{fig:ZX_Sarah} contains expressions for the $ZX$ term from the different Hamiltonian models; principle of least action, first order perturbative expression, third-order perturbative expression, and the ideal qubit limit. As expected the perturbative expressions match the principle of least action for weak $\Omega$ but diverge as $\Omega$ grows large. In addition there is a significant deviation between the $ZX$ coefficient for the perfect qubit model and that from the principle of least action which indicates the presence of higher levels with finite anharmonicity needs to be taken into account for accurate Hamiltonian modeling.

\begin{figure}[h!]
\centering
\includegraphics[width=0.43\textwidth]{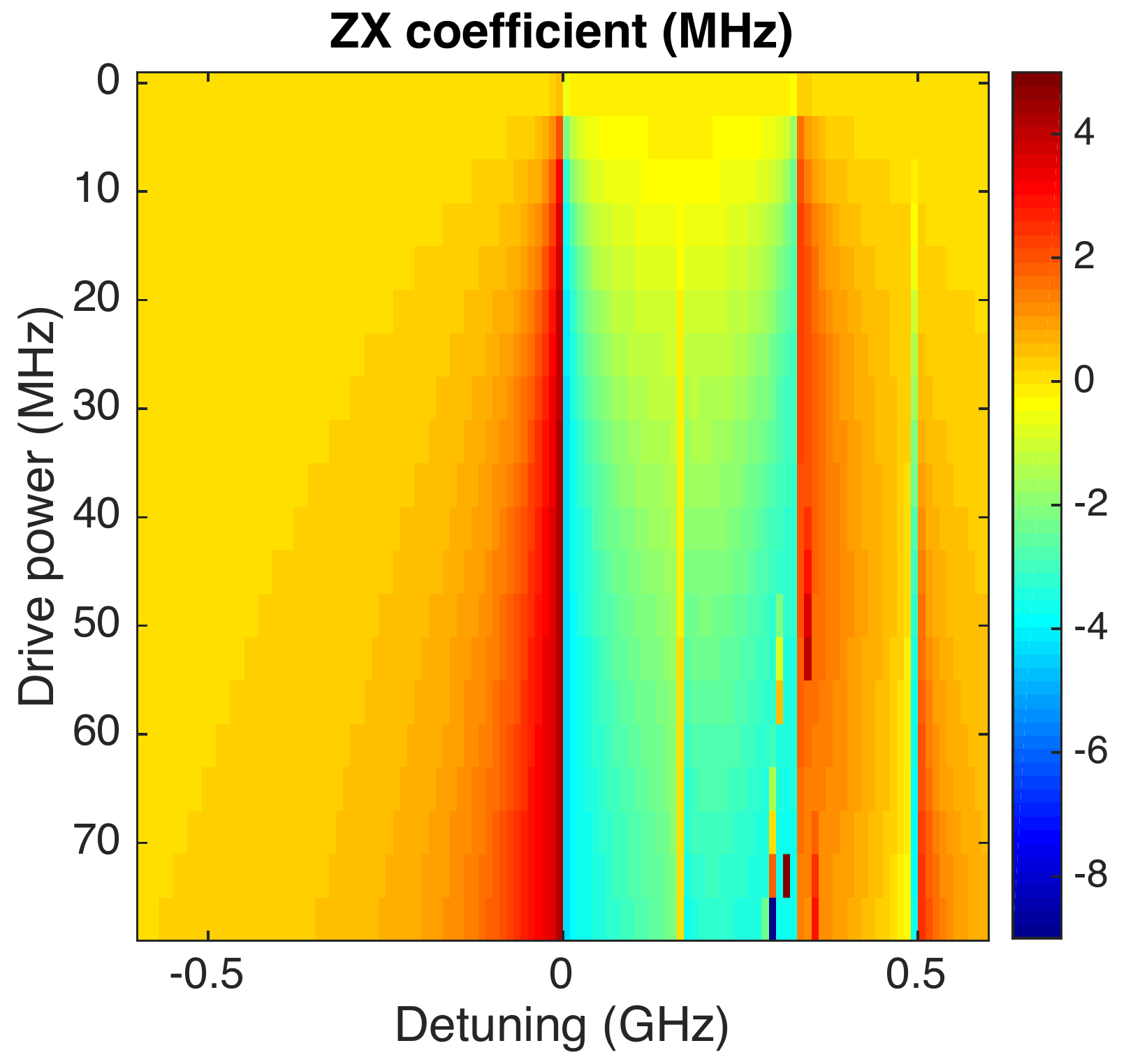}
\caption{\label{fig:ZX_2D_Sarah}  Two-dimensional $ZX$ coefficient sweep (color scale in MHz).}
\end{figure}

Next, both the frequency of the control transmon, $\omega_1$, and the drive amplitude $\Omega$ are swept with $\Delta/2\pi$ varied from 0 to 600 MHz and $\Omega/2\pi$ from 0 to 100 MHz. From the poles in the expressions of the Pauli coefficients found in Sec.~\ref{sec:higherlevelpert} one expects that when $\Delta = -\frac{\delta_1}{2}$, $-\delta_1$, $-\frac{3\delta_1}{2}$, $\Heff$ will be a poor model for $H$. A method for quantifying how well $\Heff$ captures the full dynamics is discussed in Sec.~\ref{sec:Leastaction}.
The $ZX$ coefficient is shown in Fig.~\ref{fig:ZX_2D_Sarah} and up to $-\delta_1$ there is a sizable $ZX$ rate, however past this point the rate quickly goes to 0. Intuitively this phenomenon is explained by the fact that when two transmons are detuned by an amount greater than their anharmonicity, they begin to look like harmonic oscillators with respect to each other. Therefore since entanglement can not be created between two harmonic oscillators, the $ZX$ term approaches 0 as the detuning $\Delta$ grows large.


\section{Modeling classical cross-talk from CR drive}\label{sec:Targetdrive}

In Ref.~\cite{Sheldon2016} a CR gate with the parameters of Sec.~\ref{sec:PrincipleLA} was calibrated based on the block-diagonal effective Hamiltonian in a scheme called partial Hamiltonian tomography. A large $IY$ term was found to be present but clearly the results of Sec.~\ref{sec:PrincipleLA} predict no such term can arise from from the Hamiltonian model considered to this point. One potential model for the source of this term that we investigate here is \emph{classical cross-talk} induced on the target from driving the control. 

To analyze this model we go back to Eq.~\ref{eq:Hddiag} and allow for a drive term on the target qubit whose amplitude and phase depend on the drive on the control. The total drive term then takes the form
\begin{align}\label{eq:Hddiagtwo}
\tilde{H}_d&= \Omega(t)\cos(\omega_dt+\phi_c)\tilde{B}_1 \nonumber \\
&+ A\Omega(t)\cos(\omega_dt+ \phi_t) \tilde{B}_2,
\end{align}
where $\tilde{B}_j = U^\dagger (b_j^{\dagger} + b_j)U$ for $j=1,2$, $A \leq 1 $ is a scale factor modeling the amplitude of the cross-talk term, and $\phi_t$ is the phase lag that occurs on the target. The values of these parameters depend on the form of the cross-talk channel. Since the cross-talk term corresponds directly to a rotation on the target qubit, the condition for block-diagonalization $\frac{\Omega}{\Delta_{12}} \ll 1$ is unchanged and the methods discussed here can be used to obtain an effective Hamiltonian. 

\begin{figure}[h!]
\centering
\includegraphics[width=0.43\textwidth]{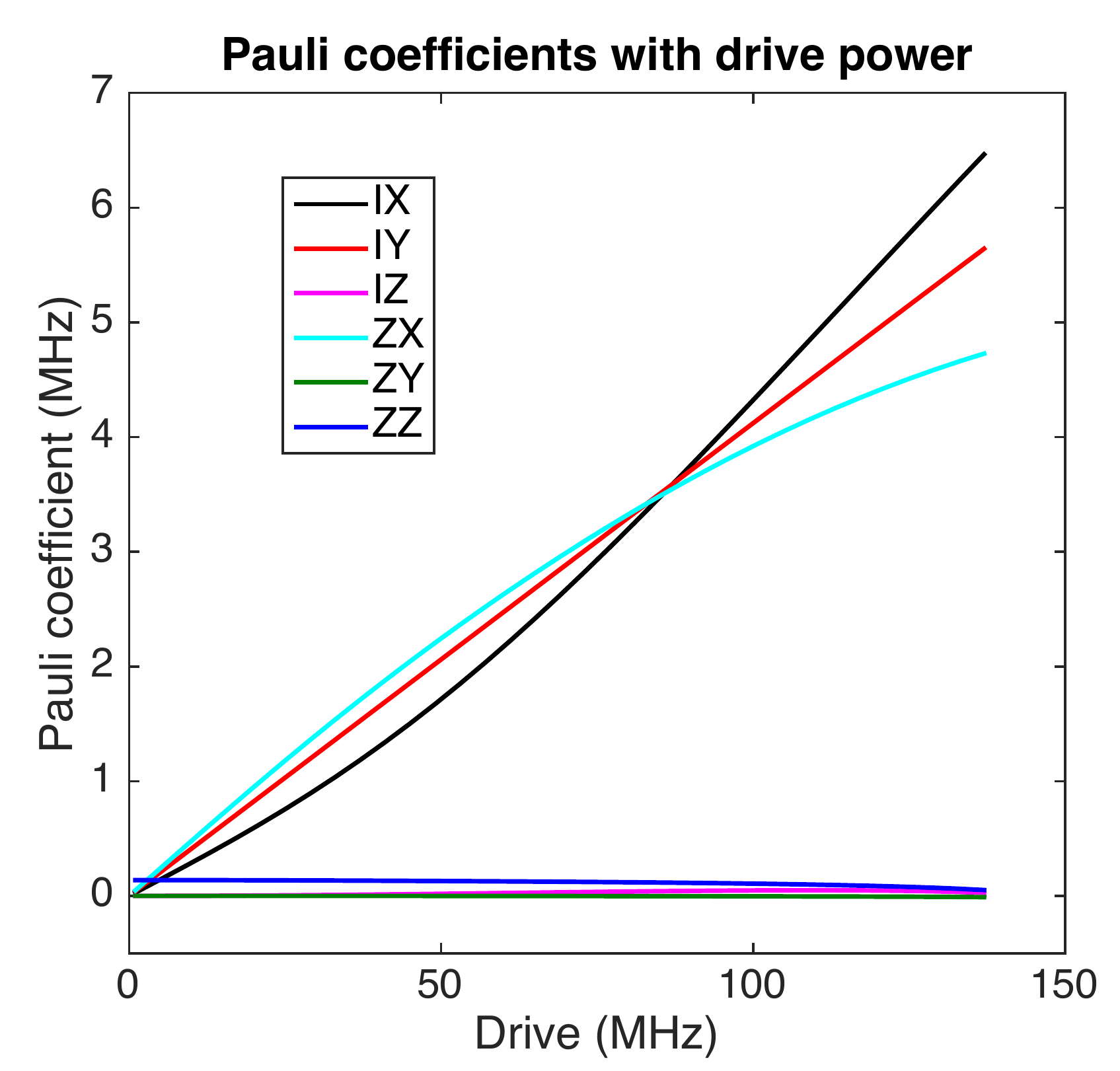}
\caption{\label{fig:Allcoeffs_Sarah_target}  Various Pauli coefficients.}
\end{figure}

Using the parameters of Ref.~\cite{Sheldon2016} we find the following values for $A$, $\phi_c$, and $\phi_t$,
\begin{align}
A&=0.071,\nonumber \\
\phi_c&=\pi,\nonumber \\
\phi_t&=-0.62,
\end{align}
produces the Pauli coefficients seen in Fig.~\ref{fig:Allcoeffs_Sarah_target} which agree well with those in Fig. 2b of Ref.~\cite{Sheldon2016}. It is important to note that this agreement only suggests classical cross-talk as a potential source for the presence of the $IY$ term in Ref.~\cite{Sheldon2016}. Potential sources of cross-talk channels are an area of current investigation.

\section{Discussion}

We have provided detailed theoretical models of the cross-resonance gate via effective block-diagonal Hamiltonian techniques. For the idealized qubit model, analytic expressions for the Hamiltonian components using the principle of least action~\cite{Cederbaum} were obtained. The only two non-zero components are a large Stark shift term on the control qubit from off-resonant driving as well as the $ZX$ term required for generating entanglement. For the realistic transmon model effective Hamiltonians were constructed via both a perturbative approach as well as the principle of least action. These two approaches agree well in the weak drive limit and predict non-zero Pauli coefficients of the form $A \otimes B$ with $A \in \{I,Z\}$, $B \in \{I,X,Z\}$. The presence of extra Hamiltonian terms compared to those from the ideal qubit case implies higher levels play an important role to understand the precise error terms for implementing a two-qubit gate.

In the experiment of Ref.~\cite{Sheldon2016} an $IY$ term was found to be present in contrast to what is predicted from our analysis with a single CR drive on the control transmon. We propose this discrepancy is a result of classical cross-talk between the two transmons and generalized the model to include this effect via an additional phase-shifted drive term on the target transmon. For a set of realistic model parameters we found good agreement between the theoretical results here and those of Ref.~\cite{Sheldon2016} which implies classical cross-talk may be a significant issue in real systems. Understanding the cross-talk channels leading to drive terms on the target transmon is an important area of further research.

Various interesting questions remain as directions for future research. First, it is useful to understand whether the perturbative construction converges to that of the principle of least action. For the standard two-block Schrieffer-Wolff transformation the perturbative construction does converge to exact unitary rotation and ideally this property holds for the multi-block case as well. It will also be interesting to apply these methods to larger multi-qubit systems, especially in the context of finding points to avoid in frequency space when dealing with fixed-frequency transmons in a circuit-QED architecture. The results from a multi-qubit analysis will have an impact on future design considerations in superconducting circuit systems.

\section{acknowledgments}

The authors thank Andrew Cross and Sarah Sheldon for helpful discussions and comments. This work was supported by the Army Research Office under contract W911NF-14-1-0124.

\begin{appendix}

\section{Effective Hamiltonians} \label{sec:EffectiveHam}

\subsection{Effective Hamiltonian from principle of least action}\label{sec:Leastaction}

Suppose one is given a Hamiltonian $H$ on the Hilbert space $\mathcal{H}$ with eigenvalues $E_a$ and eigenvectors $|s_a\rangle$;
\begin{align}\nonumber
H &= \sum_a E_a |s_a\rangle\langle s_a |.
\end{align}
A Hermitian matrix $H_\text{eff}$ is said to be an effective Hamiltonian for $H$ with respect to the orthogonal subspaces $\{\mathcal{K}_a\}$ ($\cup _a \mathcal{K}_a = \mathcal{H}$) if the following are satisfied,
\begin{enumerate}
\item $H_\text{eff}$ has the same energy spectrum as $H$,
\item $H_\text{eff}$ only has support on the $\mathcal{K}_a$.
\end{enumerate}
Suppose each subspace $\mathcal{K}_a$ has dimension $d_{\mathcal{K}_a}$ and let $P_{\mathcal{K}_a}$ be the projector onto $\mathcal{K}_a$. We set an orthonormal basis for each $\mathcal{K}_a$, denoted $\{|q^{\mathcal{K}_a}_b\rangle\}$, $b=1,...,d_{K_a}$, to be the standard basis for working in coordinates. Note that for each $\mathcal{K}_a$ any linear combination of the $|q^{\mathcal{K}_a}_b\rangle$ is still supported only on $\mathcal{K}_a$. The full orthonormal basis for $\mathcal{H}$ comprised of the union of these bases will be denoted $\{|q_a\rangle\}$. $H_\text{eff}$ is uniquely defined by a unitary matrix $T$ that maps the eigenvectors of $H$, $|s_a\rangle$, to the eigenvectors $|r_a\rangle$ of $H_\text{eff}$ with the eigenvalues being preserved since $T$ is unitary. From the desired form of $H_\text{eff}$ having support only on the $\mathcal{K}_a$, the sole restriction on the $|r_a\rangle$ is that the first $d_{\mathcal{K}_1}$ vectors have support only on $\mathcal{K}_1$, the next $d_{\mathcal{K}_2}$ have support only on $\mathcal{K}_2$, and so on. 

Let us now discuss how to actually compute $T$. The first step is to map the eigenvalues of $H$ onto the $\{|q_j\rangle\}$ basis via the eigenvector matrix $X$ of $H$ so that all of the freedom in computing $T$ comes from choosing a block-diagonal (with respect to $\{|q_j\rangle\}$) unitary matrix $F$. Since $H = \sum_a E_a |s_a\rangle\langle s_a |$, the columns of $X$ are equal to $|s_j\rangle$ when written with respect to the basis $\{|q_k\rangle\}$,
\begin{align}\label{eq:S2def}
X&=\sum_j|s_j\rangle\langle q_j|,
\end{align}
and so
\begin{align}
X^\dagger H X &= \left(\sum_j|q_j\rangle\langle s_j| \right)\left(\sum_a E_a |s_a\rangle\langle s_a |\right)\left(\sum_k|s_k\rangle\langle q_k| \right)\nonumber \\
&= \sum_a E_a |q_a\rangle\langle q_a |. \nonumber
\end{align}
The unitary matrix $F$ now rotates into the desired eigenbasis $\{|r_j\rangle\}$ and since the $|r_j\rangle$ only have support on the subspaces $\mathcal{K}_a$, $F$ represented in $|q_j\rangle$ is a unitary block-diagonal matrix. 
The total block-diagonalizing unitary $T$ can be written as the composition of $F$ with $X$ where $X$ is given in Eq.~\ref{eq:S2def} and
\begin{align}\label{eq:F2def}
F&=\sum_j|r_j\rangle\langle q_j|.
\end{align}

It is clear the freedom in choosing $H_\text{eff}$ comes entirely from choosing $F$. Ideally, one would like to obtain a unique $H_\text{eff}$ given $H$. The approach given in Ref.~\cite{Cederbaum} is to solve the following optimization problem:
\begin{align}\label{eq:argmin}
\text{argmin}_{F}\left(\|T-\mathcal{I}\|_2\right),
\end{align}
which means to find the unitary matrix $F$ that minimizes the 2-norm (Euclidean) distance between $T$ and $\mathcal{I}$. The unique solution of this problem is given by
\begin{align}
F&=\frac{X_{BD}}{\sqrt{X_{BD}X_{BD}^\dagger}},\nonumber
\end{align}
where $X_{BD}$ is the projection of $X$ onto the subspaces $\mathcal{K}_a$ and is assumed to be non-singular. Intuitively this can be thought of as first rotating $H$ into its eigenvalue matrix and attempting to rotate back to $H$ under the constraint of block-diagonality.

\begin{figure}[h!]
\centering
\includegraphics[width=0.43\textwidth]{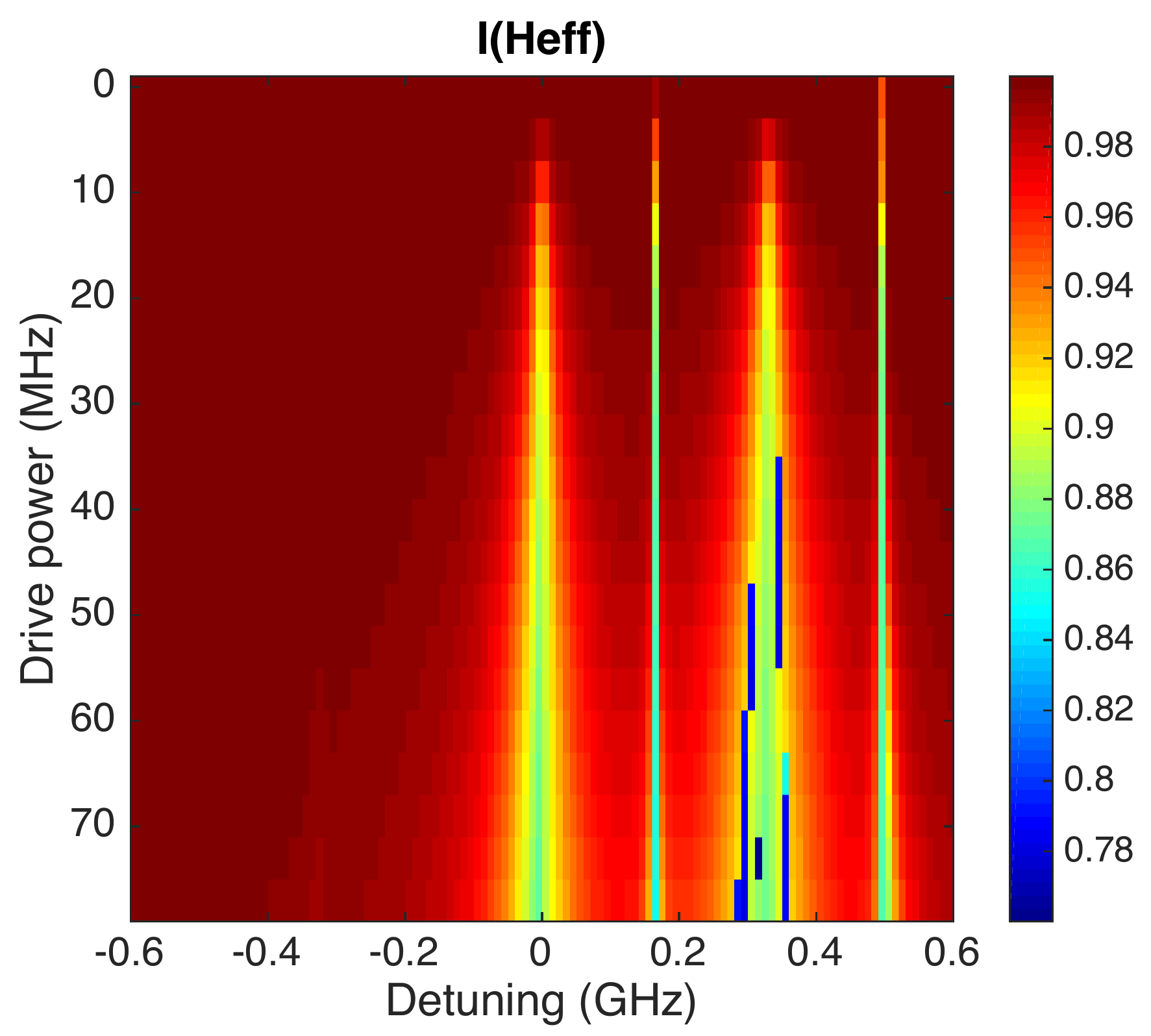}
\caption{\label{fig:IHeff_2D_Sarah}  $I(\Heff)$ with transmon detuning and drive power.}
\end{figure}

There are a variety of different metrics one could use to quantify the extent to which $H_\text{eff}$ captures the dynamics of $H$. For instance one could directly compute the objective function in Eq.~\ref{eq:argmin}. Alternatively, one can see that $H=H_\text{eff}$ if and only if $X=X_{BD}$ and if the eigenvectors of $H$ are highly mixed across different blocks then the quality of $H_\text{eff}$ as a model of $H$ decreases. As a result one can define a simple figure of merit, denoted $I(\Heff)$, to be the normalized sum of the squared magnitudes of the eigenvectors of $H$ after being projected onto the subspaces $\mathcal{K}_a$,
\begin{align}
I(\Heff) &= \frac{\text{tr}\left(X_{BD}X_{BD}^\dagger\right)}{\text{dim}(\mathcal{H})} = \frac{\|X_{BD}\|_2^2}{\text{dim}(\mathcal{H})}.\nonumber
\end{align}
Since
\begin{align}
0 \leq \text{tr}\left(X_{BD}X_{BD}^\dagger\right) \leq \text{dim}(\mathcal{H}),\nonumber
\end{align}
$I(\Heff) \in [0,1]$. A plot of $I(\Heff)$ for the parameters of Ref.~\cite{Sheldon2016} is contained in Fig.~\ref{fig:IHeff_2D_Sarah} where the control transmon frequency is fixed $\omega_1/2\pi = 5.114$ GHz. As expected $I(\Heff)$ deviates from 1 near the poles predicted from the perturbative analysis in Sec.~\ref{sec:higherlevelpert}

\subsection{Effective Hamiltonian from perturbative construction of canonical transformations} \label{sec:EffectivePerturbation}

In this section we will discuss how to perturbatively obtain a canonical transformation $U=e^{-i S}$ and an effective Hamiltonian $\Heff$ that describes the dynamics of our system. We will focus our attention on the case where we have an unperturbed Hamiltonian $H_0$ that we can solve exactly and a perturbative term $H_1$ with order parameter $\lambda$ such that
\begin{align}
H&=H_0+\lambda H_1. \nonumber
\end{align}
Obtaining $\Heff$ from $H$ depends on the desired form we would like $\Heff$ to have.  Here, we derive an iterative procedure to build the Hermitian matrix $S$ which provides the desired form of the Hamiltonian. Common examples of this method corresponds to diagonalization of $H$ and modeling the dynamics on the low-energy subspace as in the standard Schrieffer-Wolff (SW) transformation. We apply these methods to provide a complete perturbative solution to the \emph{simultaneous block-diagonalization} of $H$ into an arbitrary number of blocks. In the case of two blocks the standard SW transformation is recovered.

The Hermitian operator $S$ in $U=e^{-iS}$ can be expanded as
\begin{align}
S &=\sum_{n=1}^{\infty}\lambda^nS_n.\nonumber
\end{align}
Computing powers of $S$ results in the expressions
\begin{align}
S&= \lambda S_1 + \lambda^2 S_2 + \lambda^3 S_3 + \lambda^4 S_4 +...\nonumber \\
S^2 &=  \lambda^2 S_1^2 + \lambda^3 (S_1 S_2 + S_2S_1) + \lambda^4(S_2^2 + S_1S_3 + S_3S_1) + ...\nonumber \\
S^3 &= \lambda^3  S_1^3 + \lambda^4 (S_1^2 S_2 +S_1S_2S_1 + S_2S_1^2 ) +...\nonumber \\
S^4 &= \lambda^4 S_1^4 + ....\nonumber
\end{align}
We can expand $U=e^{\pm iS}$ in an exponential series to obtain
\begin{align}
e^{\pm i S} &= \openone \pm i(\lambda S_1 + \lambda^2 S_2 + \lambda^3 S_3 + \lambda^4 S_4 + ...) \nonumber \\
&- \frac{1}{2!} (\lambda^2 S_1^2 + \lambda^3 (S_1S_2 + S_2 S_1 + ...) \nonumber \\
&+ \lambda^4 (S_2^2 + S_1S_3 + S_3S_1 +...))\nonumber \\
& \mp \frac{i}{3!} (\lambda^3 S_1^3 + \lambda^4(S_1^2S_2 + S_1S_2S_1 + S_2S_1^2) + ...) \nonumber \\
&+ \frac{1}{4!}\left(\lambda^4 S_1^4 + ...\right) + ...\nonumber
\end{align}
Collecting in powers of $\lambda$ we have
\begin{align}
e^{\pm i S} &= \openone + \lambda(\pm i S_1) + \lambda^2 \left(\pm i S_2-\frac{1}{2}S_1^2\right) \nonumber \\
&+ \lambda^3\left(\pm i S_3 - \frac{1}{2}(S_1S_2 + S_2 S_1) \mp \frac{i}{6} S_1^3\right) \nonumber \\
&+ \lambda^4 \Bigg(\pm i S_4 - \frac{1}{2}(S_2^2 + S_1S_3 + S_3S_1) \nonumber \\
&\mp \frac{i}{6} (S_1^2S_2 + S_1S_2S_1 + S_2 S_1^2) + \frac{1}{24}S_1^4\Bigg).\label{eq:exponentials}
\end{align}
Writing
\begin{align}
e^{iS}(H_0 + \lambda V)e^{-iS} &= e^{iS}H_0e^{-iS} + \lambda e^{iS}V e^{-iS},\nonumber
\end{align}
we first deal with $e^{iS}H_0e^{-iS}$ from which an expression for $ \lambda e^{iS}V e^{-iS}$ will follow in a straightforward manner.

Expanding the exponentials in Eq.~(\ref{eq:exponentials}) and collecting powers in $\lambda$ gives the following coefficients at each order. 

\medskip

\underline{$\lambda^0$}: $H_0$.

\medskip

\underline{$\lambda^1$}: $i[S_1,H_0]$. 

\medskip

\underline{$\lambda^2$}:  $-\frac{1}{2}[S_1,[S_1,H_0]] = -\frac{S_1^2}{2}H_0 - H_0\frac{S_1^2}{2} + S_1H_0S_1$,
\begin{align}
i[S_2,H_0] &= iS_2H_0-iH_0S_2. \nonumber 
\end{align}

\underline{$\lambda^3$}: $i[S_3,H_0]$,
\begin{gather}
-\frac{i}{6}[S_1,[S_1,[S_1,H_0]]] \nonumber \\
= i\Big(\frac{3S_1^2H_0S_1}{6} - \frac{3S_1H_0S_1^2}{6} + \frac{H_0S_1^3}{6} - \frac{S_1^3H_0}{6}\Big),\nonumber 
\end{gather}
and
\begin{gather}
-\frac{1}{2}\left([S_1,[S_2,H_0]] + [S_2,[S_1,H_0]]\right) \nonumber \\
= \frac{1}{2}\Big(2S_1H_0S_2 + 2S_2H_0S_1 - H_0S_1S_2 \nonumber \\
- H_0S_2S_1 - S_1S_2H_0 - S_2S_1H_0\Big). \nonumber 
\end{gather}
\underline{$\lambda^4$}: $i[S_4,H_0]$,
\begin{gather}
-\frac{i}{6}\left([S_1,[S_1,[S_2,H_0]]] + [S_1,[S_2,[S_1,H_0]]] + [S_2,[S_1,[S_1,H_0]]]\right) \nonumber \\
= \frac{i}{6}H_0\left(S_1^2S_2 + S_1S_2S_1 + S_2S_1^2\right) \nonumber \\
+ -\frac{i}{6}\left(S_1^2S_2 + S_1S_2S_1 + S_2 S_1^2\right)H_0\nonumber \\
-\frac{i}{2}S_1H_0(S_1S_2 + S_2S_1) - \frac{i}{2}S_2H_0S_1^2 \nonumber \\
+ \frac{i}{2}S_1^2H_0S_2 + \frac{i}{2}(S_1S_2 + S_2S_1)H_0S_1, \nonumber 
\end{gather}
\begin{gather}
\frac{1}{24}[S_1,[S_1,[S_1,[S_1,H_0]]]] \nonumber \\
= \frac{1}{24}\left(H_0S_1^4 - 4S_1H_0S_1^3 + 6S_1^2H_0S_2^2 - 4S_1^3H_0S_2 + S_1^4H_0\right), \nonumber 
\end{gather}
\begin{align}
-\frac{1}{2}[S_2,[S_2,H_0]] &= -\frac{1}{2}\left(H_0S_2^2 - 2S_2H_0S_2 + S_2^2H_0\right), \nonumber 
\end{align}
\begin{gather}
-\frac{1}{2}\left([S_1,[S_3,H_0]] + [S_3,[S_1,H_0]]\right) \nonumber \\
= -\frac{1}{2}\Big(H_0(S_1S_3 + S_3S_1) - 2S_1H_0S_3 - 2S_3H_0S_1 \nonumber \\
+ (S_1S_3 + S_3S_1)H_0\Big). \nonumber 
\end{gather}
This gives to $5$'th order in $\lambda$
\begin{gather}
e^{iS}H_0e^{-iS} = H_0 + \lambda\left( i[S_1,H_0]\right) \nonumber \\
+ \lambda^2\left( i[S_2,H_0]-\frac{1}{2}[S_1,[S_1,H_0]]\right) \nonumber \\
+ \lambda^3\Big(i[S_3,H_0] -\frac{i}{6}[S_1,[S_1,[S_1,H_0]]] \nonumber \\
-\frac{1}{2}\left([S_1,[S_2,H_0]] + [S_2,[S_1,H_0]]\right)\Big) \nonumber \\
+ \lambda^4 \Bigg(i[S_4,H_0] -\frac{i}{6}\Big([S_1,[S_1,[S_2,H_0]]] \nonumber \\
 [S_1,[S_2,[S_1,H_0]]] + [S_2,[S_1,[S_1,H_0]]]\Big) \nonumber \\
 + \frac{1}{24}[S_1,[S_1,[S_1,[S_1,H_0]]]] -\frac{1}{2}[S_2,[S_2,H_0]] \nonumber \\
 -\frac{1}{2}\left([S_1,[S_3,H_0]] + [S_3,[S_1,H_0]]\right)\Bigg).\nonumber
\end{gather}
Replacing $H_0$ with $\lambda V$ we see that to $5$'th order in $\lambda$
\begin{gather}
e^{iS}\lambda Ve^{-iS} = \lambda V + \lambda^2\left( i[S_1,V]\right) \nonumber \\
+ \lambda^3\left( i[S_2,V]-\frac{1}{2}[S_1,[S_1,V]]\right) \nonumber \\
+ \lambda^4\Big(i[S_3,V] -\frac{i}{6}[S_1,[S_1,[S_1,V]]] \nonumber \\
-\frac{1}{2}\left([S_1,[S_2,V]] + [S_2,[S_1,V]]\right)\Big).\nonumber
\end{gather}
Hence in total
\begin{gather}
e^{iS}(H_0+\lambda V)e^{-iS} = H_0 + \lambda(i[S_1,H_0] + V) \nonumber \\
+ \lambda^2\left(i[S_2,H_0]  -\frac{1}{2}[S_1,[S_1,H_0]] + i[S_1,V]\right) \nonumber \\
+ \lambda^3\Big(i[S_3,H_0] -\frac{i}{6}[S_1,[S_1,[S_1,H_0]]] \nonumber \\
-\frac{1}{2}\left([S_1,[S_2,H_0]] + [S_2,[S_1,H_0]]\right) \nonumber \\
+  i[S_2,V]-\frac{1}{2}[S_1,[S_1,V]]\Big) \nonumber \\
+ \lambda^4 \Bigg(i[S_4,H_0] -\frac{i}{6}\Big([S_1,[S_1,[S_2,H_0]]] + [S_1,[S_2,[S_1,H_0]]] \nonumber \\
+ [S_2,[S_1,[S_1,H_0]]]\Big)  + \frac{1}{24}[S_1,[S_1,[S_1,[S_1,H_0]]]]\nonumber \\
-\frac{1}{2}[S_2,[S_2,H_0]] -\frac{1}{2}\left([S_1,[S_3,H_0]] + [S_3,[S_1,H_0]]\right) \nonumber \\
+ i[S_3,V] -\frac{i}{6}[S_1,[S_1,[S_1,V]]] \nonumber \\
-\frac{1}{2}\left([S_1,[S_2,V]] + [S_2,[S_1,V]]\right)\Bigg) \nonumber \\
+ O\left(\lambda^5\right).\label{eq:Totaleq}
\end{gather}
Eq.~\ref{eq:Totaleq} can be written in a more compact fashion by defining two sequences of functions $\left\{f_j = f_j\left(\{A_i\}_{i=1}^{j+1}\right)\right\}_{j=0}^\infty$ and $\left\{H^{(j)} = H^{(j)}\left(\{A_i\}_{i=1}^{j+2}\right)\right\}_{j=0}^\infty$ where the $A_i$ are indeterminate variables indicating the number of inputs to each function,
\begin{align}
e^{iS}(H_0+\lambda V)e^{-iS} &= \sum_{k=0}^\infty \lambda^k H^{(k)}\left(\{S_j\}_{j=1}^k,H_0,V\right) \nonumber \\
&=  \lambda^0 \left[f_0(H_0)\right] \nonumber \\
& +\lambda^1  \left[f_1(\{S_j\}_{j=1}^1,H_0) + f_0(V)\right] \nonumber \\
& +\lambda^2  \left[f_2(\{S_j\}_{j=1}^2,H_0) + f_1(\{S_j\}_{j=1}^1,V)\right] \nonumber \\
& +\lambda^3  \left[f_3(\{S_j\}_{j=1}^3,H_0) + f_2(\{S_j\}_{j=1}^2,V)\right] \nonumber \\
& +\lambda^4  \left[f_4(\{S_j\}_{j=1}^4,H_0) + f_3(\{S_j\}_{j=1}^3,V)\right] \nonumber \\
& + O\left(\lambda^5\right). \nonumber 
\end{align}
The $f_j$ can be constructed in a straightforward manner which allows for the computation of the perturbation to any order. First, write all decompositions of $k>0$ into a sum of non-negative integers as follows
\begin{align}
k &: \: \: (k),\nonumber \\
k-1&: \: \: (k-1,1), (1,k-1), \nonumber \\
k-2&: \: \: (k-2,1,1),(1,k-2,1),(1,1,k-2),\nonumber \\
& \: \: \: \: \: \:  (k-2,2),(2,k-2), \nonumber \\
k-3&: \: \: (k-3,1,1,1),(1,k-3,1,1),(1,1,k-3,1),\nonumber \\
& \: \: \: \: \: \:  (1,1,1,k-3),(k-3,1,2),(k-3,2,1), \nonumber \\
& \: \: \: \: \: \:  (1,k-3,2),(2,k-3,1),(1,2,k-3),(2,1,k-3), \nonumber \\
. \nonumber \\
. \nonumber \\
. \nonumber \\
0 &: \: \: (1,1,1,...,1),\nonumber 
\end{align}
where $(1,1,1,...,1)$ has $k$ indices. We now take each $(j_1,...,j_b)$ from the above expression and make the assignment
\begin{align}
(j_1,...,j_b) \rightarrow \frac{i^b}{b!}[S_{j_1},[S_{j_2},...,[S_{j_{n-1}},[S_{j_n},H_0]]...]].\nonumber 
\end{align}
As an example we compute the fifth order expression. We have
\begin{gather}
(5), \nonumber \\
(4,1), (1,4), \nonumber \\
(3,1,1),(1,3,1),(1,1,3),\nonumber \\
(3,2),(2,3),\nonumber \\
(2,1,1,1),(1,2,1,1),(1,1,2,1),(1,1,1,2),\nonumber \\
(2,2,1),(2,1,2),(1,2,2),\nonumber \\
(1,1,1,1,1).\nonumber 
\end{gather}
This gives
\begin{gather}
f_5(\{S_j\}_{j=1}^5,H_0) \nonumber \\
 = \frac{i}{1!}[S_5,H_0] \nonumber \\
 - \frac{1}{2!}\left([S_4,[S_1,H_0]] + [S_1,[S_4,H_0]]\right) \nonumber \\
 - \frac{i}{3!} \Big([S_3,[S_1,[S_1,H_0]]] + [S_1,[S_3,[S_1,H_0]]] \nonumber \\
 + [S_1,[S_1,[S_3,H_0]]] + [S_3,[S_2,H_0]] + [S_2,[S_3,H_0]]\Big) \nonumber \\
+ \frac{1}{4!}\Big( [S_2,[S_1,[S_1,[S_1,H_0]]]] +  [S_1,[S_2,[S_1,[S_1,H_0]]]] \nonumber \\
+  [S_1,[S_1,[S_2,[S_1,H_0]]]] +  [S_1,[S_1,[S_1,[S_2,H_0]]]] \nonumber \\
+ [S_2,[S_2,[S_1,H_0]]] + [S_2,[S_1,[S_2,H_0]]] + [S_1,[S_2,[S_2,H_0]]]\Big) \nonumber \\
+ \frac{i}{5!} [S_1,[S_1,[S_1,[S_1,[S_1,H_0]]]]].\nonumber 
\end{gather}
Combining this with the expression for $f_4\left(\{S_j\}_{j=1}^4,V\right)$ already computed gives the full fifth order term
\begin{gather}
\lambda^5 \Bigg[\frac{i}{1!}[S_5,H_0] - \frac{1}{2!}\left([S_4,[S_1,H_0]] + [S_1,[S_4,H_0]]\right) \nonumber \\
 - \frac{i}{3!} \Big([S_3,[S_1,[S_1,H_0]]] + [S_1,[S_3,[S_1,H_0]]] \nonumber \\
 + [S_1,[S_1,[S_3,H_0]]]  + [S_3,[S_2,H_0]] + [S_2,[S_3,H_0]]\Big) \nonumber \\
+ \frac{1}{4!}\Big( [S_2,[S_1,[S_1,[S_1,H_0]]]] +  [S_1,[S_2,[S_1,[S_1,H_0]]]] \nonumber \\
+  [S_1,[S_1,[S_2,[S_1,H_0]]]] +  [S_1,[S_1,[S_1,[S_2,H_0]]]] \nonumber \\
+ [S_2,[S_2,[S_1,H_0]]] + [S_2,[S_1,[S_2,H_0]]] \nonumber \\
+ [S_1,[S_2,[S_2,H_0]]]\Big) + \frac{i}{5!} [S_1,[S_1,[S_1,[S_1,[S_1,H_0]]]]]  \nonumber \\
+ \Bigg(i[S_4,V] -\frac{i}{6}\Big([S_1,[S_1,[S_2,V]]] + [S_1,[S_2,[S_1,V]]] \nonumber \\
+ [S_2,[S_1,[S_1,V]]]\Big) + \frac{1}{24}[S_1,[S_1,[S_1,[S_1,V]]]]\nonumber \\
-\frac{1}{2}[S_2,[S_2,V]] -\frac{1}{2}\left([S_1,[S_3,V]] + [S_3,[S_1,V]]\right)\Bigg)\Bigg].\nonumber 
\end{gather}

Now that we can compute each $f_j$ we are able to recursively compute every order $H^{(j)}$. What remains is to compute the $S_j$ which is done by noting that at each order, $H^{(k)}\left(\{S_j\}_{j=1}^k,H_0,V\right)$ contains only one term with $S_k$ in it, $i[S_k,H_0]$. Hence one can write
\begin{gather}\label{eq:Hm}
H^{(m)}\left(\{S_j\}_{j=1}^m,H_0,V\right) \nonumber \\
= i[S_m,H_0] + H_x^{(m)}\left(\{S_j\}_{j=1}^{m-1},H_0,V\right),\label{eq:Hm}
\end{gather}
and assuming $\{S_j\}_{j=1}^{k-1}$ have already been computed, $H_x^{(k)}$ can be computed as well. Hence one need only solve for $S_k$ at each order to compute $H^{(k)}$. $S_k$ is computed by ensuring $H^{(k)}$ satisfies the required form set by the problem. We now illustrate the method with two examples, diagonalization and block-diagonalization.

\subsubsection{Example 1: Diagonalization} \label{sec:Diagonalization}

Suppose we want our effective dynamics to be diagonal at each order $m$, that is we want $H^{(m)}$ to be diagonal for every $m$ ($H_0$ is diagonal and $V$ is a perturbation containing off-diagonal components). We have 
\begin{align}
H^{(0)} &= H_0, \nonumber \\
H_x^{(1)} &= H_1.\nonumber
\end{align}
One can see from Eq.~(\ref{eq:Hm}) that if $H^{(m)}$ is diagonal
\begin{align}
\sum_pE_p^{(m)} |p\rangle\langle p| &= i \sum_pE_p^{(0)}(S^{(m)}|p\rangle\langle p| - |p\rangle\langle p|S^{(m)}) \nonumber \\
& + H_x^{(m)}.
\end{align}
Without loss of generality we can assume that $S$ is an off-diagonal matrix (has diagonal entries of $0$) and so the above is satisfied if
\begin{align}
E_p^{(m)} &= \langle p|H_x^{(m)} |p\rangle, \nonumber \\
\langle p|S^{(m)}|q\rangle &= \frac{-i\langle p |H_x^{(m)} |q\rangle}{E_p^{(0)} - E_q^{(0)}}, p \neq q.\nonumber
\end{align}

\subsubsection{Example 2: Block-Diagonalization} \label{sec:Blockdiagonalization}

Suppose we want our effective dynamics to be block-diagonal at each order $m$, that is we want $H^{(m)}$ to be block-diagonal for every $m$ ($H_0$ is block-diagonal and $V$ is a perturbation containing off-block-diagonal components). We have 
\begin{align}
H^{(0)} &= H_0, \nonumber \\
H_x^{(1)} &= H_1.\nonumber
\end{align}
One can see from Eq.~(\ref{eq:Hm}) that if $H^{(m)}$ is block-diagonal
\begin{align}
H^{(m)} & = H_1^{(m)} \oplus ... \oplus H_k^{(m)} \oplus ... \nonumber
\end{align}
then acting subspace projectors $P_j$ and $P_k$ on both sides of Eq.~(\ref{eq:Hm}) gives
\begin{gather}
P_j  H_1^{(m)} \oplus ... \oplus H_k^{(m)} \oplus ... P_k \nonumber \\
= i \Big(P_j S^{(m)} ( H_1^{(0)} \oplus ... \oplus H_k^{(0)} \oplus ...) P_k \nonumber \\
- P_j  ( H_1^{(0)} \oplus ... \oplus H_k^{(0)} \oplus ...) S^{(m)}\Big) P_k + P_j H_x^{(m)} P_k, \nonumber
\end{gather}
\begin{gather}
H_j^{(m)}\delta_{j,k} = i(P_j S^{(m)} P_k H_k^{(0)} - H_j^{(0)} P_j S^{(m)} P_k) + H_{x_{j,k}}^{(m)}, \nonumber
\end{gather}
and
\begin{gather}
i H_{x_{j,k}}^{(m)} + H_j^{(m)} \delta_{j,k} = S_{j,k}^{(m)}H_k^{(0)} - H_j^{(0)} S_{j,k}^{(m)}. \nonumber
\end{gather}
Since $S$ is an off-block-diagonal matrix (has block-diagonal entries of $0$) we have
\begin{align}
H_j^{(m)} &=  H_{x_{j,j}}^{(m)}, \text{if} \: j=k, \nonumber \\
H_j^{(0)} S_{j,k}^{(m)} - S_{j,k}^{(m)}H_k^{(0)}  &= - i H_{x_{j,k}}^{(m)},  \text{if } \:  j \neq k. \nonumber
\end{align}

In the case that $H^{(0)}$ is diagonal we can solve easily for $S^{(m)}$ at each order,
\begin{align}
\langle p|S_{j,k}^{(m)}|q\rangle &= \frac{-i\langle p |H_{x_{j,k}}^{(m)} |q\rangle}{\langle p| H_j^{(0)} |p \rangle - \langle q |H_k^{(0)} |q\rangle}. \nonumber
\end{align}
However if $H^{(0)}$ is not diagonal we need to use the following matrix-vector correspondence. For any $A$, $B$, $C$
\begin{align}
(A\otimes B)\text{vec}(C) &= \text{vec}(ACB^T), \nonumber
\end{align}
where ``$\text{vec}$" is defined as $\text{vec}(|a\rangle \langle b|) = |a\rangle \otimes |b\rangle$.
Hence
\begin{align}
AB-BC = D\nonumber &\Leftrightarrow \nonumber \\
AB\openone^T - \openone BC = D &\Leftrightarrow \nonumber \\
\text{vec}(AB\openone^T) - \text{vec}( \openone BC) = \text{vec}(D) &\Leftrightarrow \nonumber \\
(A\otimes \openone - \openone \otimes C^T) \text{vec}(B) = \text{vec}(D) &\Leftrightarrow \nonumber \\
\text{vec}(B) = (A\otimes \openone - \openone \otimes C^T)^{-1}\text{vec}(D) &\Leftrightarrow \nonumber \\
B = \text{mat}\left(  (A\otimes \openone - \openone \otimes C^T)^{-1}\text{vec}(D)\right). \nonumber
\end{align}
Letting
\begin{align}
A&= H_j^{(0)} \nonumber \\
B&= S_{j,k}^{(m)}\nonumber \\
C&= H_k^{(0)} \nonumber \\
D&=  - i H_{x_{j,k}}^{(m)}, \nonumber
\end{align}
allows for $S_{j,k}^{(m)}$ to be solved at each order $m$.

\subsubsection{Summary of results for perturbative construction}

The main result is
\begin{align}
\Heff &= U^\dagger H U = \sum_{m=0}^\infty \lambda^m H^{(m)}, \nonumber
\end{align}
where
\begin{align}
H^{(m)} &= H^{(m)}\left(\{S_j\}_{j=1}^m,H_0,H_1\right) \nonumber \\
&= i[S_m,H_0] + H_x^{(m)}\left(\{S_j\}_{j=1}^{m-1},H_0,H_1\right). \nonumber
\end{align}
At each order $m$, $H_x^{(m)}$ is a function of only $\{S_1,...,S_{m-1}\}$ and so can be computed since we assume the lower order $\{S_1,...,S_{m-1}\}$ are known. Hence $H^{(m)}\left(\{S_j\}_{j=1}^m,H_0,H_1\right)$ has only one term containing $S_m$, $i[S_m,H_0]$. Solving for $S_m$ subject to the desired dynamics allows for computation of $H^{(m)}$. For instance in the case of the SW transformation, the desired dynamics is to have $H^{(m)}$ be block-diagonal on $H_l$ and $H_e$. 

We have
\begin{align}
H^{(0)} &= H_0, \nonumber \\
H_x^{(1)} &= H_1. \nonumber
\end{align}
If $H^{(m)}$ is block-diagonal
\begin{align}
H^{(m)} & = H_1^{(m)} \oplus ... \oplus H_k^{(m)} \oplus ... \nonumber
\end{align}
then since $S$ can without loss of generality be an off-block-diagonal matrix (the block-diagonal entries of $S$ are $0$)
\begin{align}
H_j^{(m)} &=  H_{x_{j,j}}^{(m)}, \text{if} \: j=k, \nonumber \\
H_j^{(0)} S_{j,k}^{(m)} - S_{j,k}^{(m)}H_k^{(0)}  &= - i H_{x_{j,k}}^{(m)},  \text{if } \:  j \neq k. \nonumber
\end{align}
Note that if $H^{(0)}$ is diagonal then
\begin{align}
\langle p|S_{j,k}^{(m)}|q\rangle &= \frac{-i\langle p |H_{x_{j,k}}^{(m)} |q\rangle}{\langle p| H_j^{(0)} |p \rangle - \langle q |H_k^{(0)} |q\rangle}. \nonumber
\end{align}
If $H^{(0)}$ is not diagonal we use the following matrix-vector correspondence
\begin{align}
(A\otimes B)\text{vec}(C) &= \text{vec}(ACB^T),\nonumber
\end{align}
which holds for any $A$, $B$, $C$, where ``$\text{vec}$" is defined as $\text{vec}(|a\rangle \langle b|) = |a\rangle \otimes |b\rangle$.
Hence
\begin{align}
AB-BC= D\nonumber &\Leftrightarrow B = \text{mat}\left(  (A\otimes \openone - \openone \otimes C^T)^{-1}\text{vec}(D)\right). \nonumber
\end{align}
Letting $A= H_j^{(0)}$, $B= S_{j,k}^{(m)}$, $C= H_k^{(0)}$, and $D=  - i H_{x_{j,k}}^{(m)}$ allows us to solve for $S_{j,k}^{(m)}$ at each order $m$.

\section{Making the RWA in the Duffing model case}\label{sec:RWADuffing}

We move into a frame rotating at $\omega_d$ on both qubits. The unitary operator $R$ corresponding to this frame transformation is defined by the Hamiltonian $\tilde{H}_A = \omega_d(b_1^\dagger b_1 +  b_2^\dagger b_2)$,
\begin{align}
R&=e^{-i\left[ \omega_d(b_1^\dagger b_1 +  b_2^\dagger b_2)\right]t}. \nonumber
\end{align}
This gives the Hamiltonian
\begin{align}
\tilde{H}_\text{sys}^{(0)}-\tilde{H}_A + R^\dagger \tilde{H}_d R &=: \tilde{H}_\text{drift} + R^\dagger \tilde{H}_d R. \nonumber
\end{align}
Let us now focus on the term $R^\dagger \tilde{H}_d R$ and make the RWA, which amounts to ignoring all excitations of energy cost $2\omega_d$ or higher.

We have
\begin{align}
&R^\dagger \tilde{H}_d R \nonumber \\
&= \sum_{j=1}^2\left[\Omega_{X_j}(t)\cos(\omega_dt) + \Omega_{Y_j}(t)\sin(\omega_dt)\right]R^\dagger \tilde{B}_j R \nonumber \\
&= \sum_{j=1}^2\Big[\Omega_{X_j}(t)\left(\frac{e^{i\omega_dt} + e^{-i\omega_dt}}{2}\right) - \nonumber \\
& \: \: \: \: \: \: \: \: \: \: \: \: \: \: \: \:  i\Omega_{Y_j}(t)\left(\frac{e^{i\omega_dt} -  e^{-i\omega_dt}}{2}\right)\Big]R^\dagger \tilde{B}_j R.\nonumber
\end{align}
First, let's analyze the term $R^\dagger \tilde{B}_1 R$,
\begin{align}
R^\dagger \tilde{B}_1 R &= e^{-i\omega_d(b_1^\dagger b_1 + b_2^\dagger b_2)t} \tilde{B}_1 e^{i\omega_d(b_1^\dagger b_1 + b_2^\dagger b_2)t}. \nonumber
\end{align}
Let
\begin{align}
\tilde{B}_1 &= \sum_{i_1,i_2,j_1,j_2}\tilde{B}_1^{i_1,i_2,j_1,j_2}|i_1d+i_2\rangle \langle j_1d + j_2|, \nonumber
\end{align}
and
\begin{align}
e^{-i\omega_d(b_1^\dagger b_1 + b_2^\dagger b_2)t} &= \sum_{i_1,i_2}e^{-i\omega_dt(i_1+i_2)}|i_1d + i_2\rangle\langle i_1d+i_2|, \nonumber
\end{align}
where the index in each sum is taken from $0$ to $d-1$. If $\vec{i}=(i_1,i_2)$, $\vec{j}=(j_1,j_2)$ then
\begin{align}
R^\dagger \tilde{B}_1 R &= \sum_{\vec{i},\vec{j}}e^{-i\omega_dt(i_1+i_2-j_1-j_2)}\tilde{B}_1^{\vec{i},\vec{j}}|i_1d+i_2\rangle \langle j_1d + j_2|. \nonumber
\end{align}
and so if 
\begin{gather}
\Omega_j :=\Omega_{X_j}(t)\left(\frac{e^{i\omega_dt} + e^{-i\omega_dt}}{2}\right) -i\Omega_{Y_j}(t)\left(\frac{e^{i\omega_dt} -  e^{-i\omega_dt}}{2}\right),\nonumber 
\end{gather} 
then
\begin{gather}
R^\dagger \tilde{H}_d R \nonumber \\
=\Omega_1\sum_{\vec{i},\vec{j}}e^{-i\omega_dt(i_1+i_2-j_1-j_2)}\tilde{B}_1^{\vec{i},\vec{j}}|i_1d+i_2\rangle \langle j_1d + j_2| \nonumber \\
+ \Omega_2\sum_{\vec{i},\vec{j}}e^{-i\omega_dt(i_1+i_2-j_1-j_2)}\tilde{B}_2^{\vec{i},\vec{j}}|i_1d+i_2\rangle \langle j_1d + j_2|. \nonumber
\end{gather}
Now we want to ignore all terms rotating at $2\omega_2$ or higher. Let us focus on the $\Omega_{X_1}(t)$ term first. We  have
\begin{gather}
\Omega_{X_1}(t)\frac{e^{i\omega_dt} + e^{-i\omega_dt}}{2}e^{-i\omega_dt(i_1+i_2-j_1-j_2)}\tilde{B}_1^{\vec{i},\vec{j}}\nonumber \\
=\frac{\Omega_{X_1}(t)}{2}\tilde{B}_1^{\vec{i},\vec{j}}\Big[e^{i\omega_dt(1-i_1-i_2+j_1+j_2)} \nonumber \\
\: \: \: \: \: \: \: \: \: \: \: \: \: \: \: \: \: \: \: \: \: \: \: \: \: \: \: \: \: \: \: \: \: \: \: \: + e^{-i\omega_dt(1+i_1+i_2-j_1-j_2)}\Big],\nonumber 
\end{gather}
\begin{gather}
-i\Omega_{Y_1}(t)\frac{e^{i\omega_dt} + e^{-i\omega_dt}}{2}e^{-i\omega_dt(i_1+i_2-j_1-j_2)}\tilde{B}_1^{\vec{i},\vec{j}}\nonumber \\ 
=\frac{-i\Omega_{Y_1}(t)}{2}\tilde{B}_1^{\vec{i},\vec{j}}\Big[e^{i\omega_dt(1-i_1-i_2+j_1+j_2)} \nonumber \\
\: \: \: \: \: \: \: \: \: \: \: \: \: \: \: \: \: \: \: \: \: \: \: \: \: \: \: \: \: \: \: \: \: \: \: \:  - e^{-i\omega_dt(1+i_1+i_2-j_1-j_2)}\Big], \nonumber 
\end{gather}
\begin{gather}
\Omega_{X_2}(t)\frac{e^{i\omega_dt} + e^{-i\omega_dt}}{2}e^{-i\omega_dt(i_1+i_2-j_1-j_2)}\tilde{B}_2^{\vec{i},\vec{j}}\nonumber \\ 
=\frac{\Omega_{X_2}(t)}{2}\tilde{B}_2^{\vec{i},\vec{j}}\Big[e^{i\omega_dt(1-i_1-i_2+j_1+j_2)} \nonumber \\
\: \: \: \: \: \: \: \: \: \: \: \: \: \: \: \: \: \: \: \: \: \: \: \: \: \: \: \: \: \: \: \: \: \: \: \: + e^{-i\omega_dt(1+i_1+i_2-j_1-j_2)}\Big], \nonumber 
\end{gather}
\begin{gather}
-i\Omega_{Y_2}(t)\frac{e^{i\omega_dt} + e^{-i\omega_dt}}{2}e^{-i\omega_dt(i_1+i_2-j_1-j_2)}\tilde{B}_2^{\vec{i},\vec{j}}\nonumber \\ 
=\frac{-i\Omega_{Y_2}(t)}{2}\tilde{B}_2^{\vec{i},\vec{j}}\Big[e^{i\omega_dt(1-i_1-i_2+j_1+j_2)} \nonumber \\
\: \: \: \: \: \: \: \: \: \: \: \: \: \: \: \: \: \: \: \: \: \: \: \: \: \: \: \: \: \: \: \: \: \: \: \:  - e^{-i\omega_dt(1+i_1+i_2-j_1-j_2)}\Big]. \nonumber
\end{gather}
Looking at the terms in the brackets on the right-hand side we see that 
\[ \begin{cases}
\mbox{if } |i_1+i_2-j_1-j_2| \neq 1 \: \:   \text{then}  \nonumber \\
  e^{i\omega_dt(1-i_1-i_2+j_1+j_2)} = 0, e^{-i\omega_dt(1+i_1+i_2-j_1-j_2)} = 0, \nonumber \\
\mbox{if } i_1+i_2-j_1-j_2 = 1  \: \:  \text{then} \nonumber \\
 e^{i\omega_dt(1-i_1-i_2+j_1+j_2)} = 1, e^{-i\omega_dt(1+i_1+i_2-j_1-j_2)} = 0,  \nonumber \\
\mbox{if } i_1+i_2-j_1-j_2 = -1  \: \: \text{then} \nonumber \\
 e^{i\omega_dt(1-i_1-i_2+j_1+j_2)} = 0, e^{-i\omega_dt(1+i_1+i_2-j_1-j_2)} = 1.  \nonumber\\
\end{cases}
\]
Hence, in total,
\begin{equation}\label{eq:Cases}
\begin{cases}
\mbox{if } |i_1+i_2-j_1-j_2| \neq 1 \:  \text{then}    (RH_\text{d,diag}R^{\dagger})_{i_1,i_2,j_1,j_2}=0 , \\
\mbox{if } i_1+i_2-j_1-j_2 = 1  \: \text{then} \\         
(R\tilde{H}_dR^{\dagger})_{i_1,i_2,j_1,j_2} =
     \frac{\Omega_{X_1}(t)-i\Omega_{Y_1}(t)}{2}\tilde{B}_1^{\vec{i},\vec{j}} \\
\: \: \: \: \: \: \: \: \: \: \: \: \: \: \: \: \: \: \: \: \: \: \: \: \: \: \: \: \: \: \: \: \: \: \: \: \: \: \: \: \: \: \: \: \: \:          +    \frac{\Omega_{X_2}(t)-i\Omega_{Y_2}(t)}{2}\tilde{B}_2^{\vec{i},\vec{j}}   ,              \\
\mbox{if } i_1+i_2-j_1-j_2 = -1  \: \text{then} \nonumber \\
(R\tilde{H}_d R^{\dagger})_{i_1,i_2,j_1,j_2}= \frac{\Omega_{X_1}(t)+i\Omega_{Y_1}(t)}{2}\tilde{B}_1^{\vec{i},\vec{j}} \\
\: \: \: \: \: \: \: \: \: \: \: \: \: \: \: \: \: \: \: \: \: \: \: \: \: \: \: \: \: \: \: \: \: \: \: \: \: \: \: \: \: \: \: \: \: \:          +    \frac{\Omega_{X_2}(t)+i\Omega_{Y_2}(t)}{2}\tilde{B}_2^{\vec{i},\vec{j}}.  \\
\end{cases}
\end{equation}
The full rotating wave approximated Hamiltonian becomes
\begin{align}
H_\text{RWA}&=\tilde{H}_\text{drift}+\tilde{H}_\text{d,RWA},\nonumber
\end{align}
where
\begin{align}
\tilde{H}_\text{drift}&:=\tilde{H}_\text{sys}^{(0)}-\tilde{H}_A, \nonumber \\
\tilde{H}_\text{d,RWA}&:= (R^\dagger \tilde{H}_d R)^\text{RWA}, \nonumber \\
\tilde{H}_A &= \omega_d(b_1^\dagger b_1 +  b_2^\dagger b_2),
\end{align}
and the matrix elements of $ (R^\dagger H_\text{d,diag}R)^\text{rwa}$ are as defined by the above cases.

\section{Pauli coefficients}\label{sec:Paulicoeff}

The full set of Pauli coefficients of the form $A \otimes B$ for $A \in \{I,Z\}$ and $B \in \{I,X,Y,Z\}$ are given below.
\begin{align}
\frac{IX}{2}_\text{coeff} &=-\frac{J
   \Omega}{\Delta+\delta_1}
 + \frac{\Delta \delta_1 J \Omega^3}{(\Delta+\delta_1)^3 (2
   \Delta+\delta_1) (2 \Delta+3 \delta_1)},\nonumber 
\end{align}
\begin{align}
   \frac{IY}{2}_\text{coeff} &=0,\nonumber
\end{align}   
\begin{align}
   \frac{IZ}{2}_\text{coeff} &= \frac{J^2 \Omega^2}{2} \Bigg(\frac{\delta_1^3-2 \delta_1
   \Delta^2-2 \Delta^3}{\delta_1 \Delta^2
   (\delta_1+\Delta)^2 (\Delta-\delta_2)}+\frac{\delta_1^2+\Delta^2}{\Delta^2
   \delta_2 (\delta_1+\Delta)^2}\nonumber \\
   &+\frac{6 \text{$\delta
   $1}^5+4 \delta_1^4 \Delta-6 \delta_1^3 \Delta^2+7 \delta_1^2 \Delta^3+12 \delta_1 \Delta^4+4 \Delta^5}{\Delta^2 (\delta_1+\Delta)^2 (2 \delta_1+\Delta)^2 (\delta_1+2 \Delta) (3 \delta_1+2 \Delta)}\nonumber \\
   &+\frac{2}{\delta_1
   (\delta_1+\Delta) (\delta_1+\Delta-\delta_2)} \nonumber \\
   &+\frac{2}{(\delta_1+\Delta)
   (\delta_1+\Delta-\delta_2)^2}+\frac{1}{\Delta (\Delta-\delta_2)^2}\Bigg),\nonumber 
\end{align}   
\begin{align}
   \frac{ZI}{2}_\text{coeff} &=-\frac{\delta_1 \Omega ^2}{2\Delta (\delta_1+\Delta)} \nonumber \\
   &+ \frac{J^2 \Omega ^2}{2
   (\delta_1+\Delta)^3} \Bigg(\frac{2 \left(\delta_1^2+\delta_1
   \Delta+\Delta^2\right) (\delta_1+\Delta)}{\delta_1 \Delta (\delta_2-\Delta)}\nonumber \\
   &+\frac{1}{2} \delta_1 \Big(\frac{4 \delta_1^2}{\Delta^3}+\frac{11 \delta_1}{\Delta^2}+\frac{3 \delta_1}{(2 \delta_1+\Delta)^2} \nonumber \\
   &- \frac{2}{\delta_1+2
   \Delta}-\frac{6}{3 \delta_1+2 \Delta}+\frac{12}{\Delta}\Big)\nonumber \\
   &+\frac{2 (\delta_1+\Delta)^2}{\delta_1 (\delta_1+\Delta-\delta_2)}+\frac{2 (\delta_1+\Delta)^2}{(\delta_1+\Delta-\delta_2)^2}-\frac{2 \delta_1 (\delta_1+\Delta)}{\Delta \delta_2}\Bigg),\nonumber 
\end{align}
\begin{align}
   \frac{ZX}{2}_\text{coeff} &= -\frac{J\Omega}{\Delta}\left(\frac{\delta_1}{\delta_1+\Delta}\right) \nonumber \\
   &+ \frac{ J\Omega^3 \delta_1^2 (3 \delta_1^3 + 11 \delta_1^2 \Delta + 15 \delta_1 \Delta^2 + 9 \Delta^3)
}{ 2\Delta^3 (\delta1 + \Delta)^3 (\delta_1 + 2 \Delta)(3 \delta_1 + 2 \Delta)},\nonumber
\end{align}
\begin{align}
      \frac{ZY}{2}_\text{coeff} &=0,\nonumber
\end{align}
\begin{align}
      \frac{ZZ}{2}_\text{coeff} &=\frac{J^2}{2 (\delta_1+\Delta)^2} \Bigg(\Omega^2 \Bigg(\frac{\delta_1^3-2 \delta_1
   \Delta^2-2 \Delta^3}{\delta_1 \Delta^2
   (\delta_2-\Delta)} \nonumber \\
   &+\frac{1}{2} \left(\frac{4 (3 \delta_1+\Delta) \left(\delta_1^2+\delta_1 \Delta+\Delta^2\right)}{\Delta^2 (2 \delta_1+\Delta)^2}-\frac{16 \Delta}{3 \delta_1^2+8
   \delta_1 \Delta+4 \Delta^2}\right)\nonumber \\
   &+\frac{2
   \delta_1}{\Delta \delta_2}-\frac{2 (\delta_1+\Delta)}{(\delta_1+\Delta-\delta_2)^2}-\frac{2 (\delta_1+\Delta)}{\delta_1
   (\delta_1+\Delta-\delta_2)}\Bigg) \nonumber \\
   &+\frac{2
   (\delta_1+\Delta) (\delta_1+\delta_2)}{\Delta-\delta_2}\Bigg).\nonumber
\end{align}

\end{appendix}


\begin{thebibliography}{10}

\bibitem{Dennis2002}
E.~Dennis et~al., Journal of Mathematical Physics \textbf{43}, 4452 (2002).

\bibitem{Harrington2007}
R.~Raussendorf and J.~Harrington, Phys. Rev. Lett. \textbf{98}, 190504 (2007).

\bibitem{Paraoanu}
G.~S. Paraoanu, Phys. Rev. B \textbf{74}, 140504 (2006).

\bibitem{Rigetti10}
C.~Rigetti and M.~Devoret, Phys. Rev. B \textbf{81}, 134507 (2010).

\bibitem{Gambetta2017}
J.~M. Gambetta, J.~M. Chow, and M.~Steffen, Nat. Phys. JQI \textbf{3}, 2
  (2017).

\bibitem{Sheldon2016}
S.~Sheldon et~al., Phys. Rev. A \textbf{93}, 060302 (2016).

\bibitem{Chow2014}
J.~M. Chow et~al., Nat. Comm. \textbf{5}, 4015 (2014).

\bibitem{Corcoles2015}
A.~D. C\'orcoles et~al., Nat. Comm. \textbf{6}, 6979 (2015).

\bibitem{Takita2016}
M.~Takita et~al., Phys. Rev. Lett. \textbf{117}, 210505 (2016).

\bibitem{Takita2017}
M.~Takita et~al., Phys. Rev. Lett. \textbf{119}, 180501 (2017).

\bibitem{SW}
J.~R. Schrieffer and P.~A. Wolff, Phys. Rev. \textbf{149}, 491 (1966).

\bibitem{Bravyi11}
S.~Bravyi, D.~DiVincenzo, and D.~Loss, Ann. Phys. \textbf{326}, 2793 (2011).

\bibitem{BornOpp}
M.~Born and R.~Oppenheimer, Annalen der Physik \textbf{389}, 457 (1927).

\bibitem{Cederbaum}
L.~S. Cederbaum, J.~Schirmer, and H.~D. Meyer, Journal of Physics A:
  Mathematical and General \textbf{22}, 2427 (1989).

\bibitem{BHW04}
A.~Blais et~al., Phys. Rev. A \textbf{69}, 062320 (2004).

\bibitem{Koch07}
J.~Koch et~al., Phys. Rev. A \textbf{76}, 042319 (2007).

\end{thebibliography}

\end{document}